\documentclass[a4paper]{article}
\usepackage{amsmath}
\usepackage{amsfonts}
\usepackage{float}
\usepackage{bbm}
\usepackage{bm}
\usepackage{listings}
\usepackage{graphicx}
\usepackage{color}

\usepackage{caption}
\usepackage{subcaption}
\usepackage{ulem}
\usepackage{algorithm}
\usepackage[noend]{algpseudocode}

\usepackage{natbib}

\makeatletter
\def\BState{\State\hskip-\ALG@thistlm}
\makeatother

\DeclareMathOperator*{\Ex}{E}
\DeclareMathOperator*{\Var}{Var}

\DeclareMathOperator*{\MAR}{MAR}

\providecommand{\keyword}[1]{\textbf{\textit{Keywords: }} #1}

\title{Bayesian analysis of mixture autoregressive models
  covering the complete parameter space
}

\author{Davide Ravagli \\ \normalsize davide.ravagli@manchester.ac.uk
	\and Georgi N. Boshnakov \\ \normalsize georgi.boshnakov@manchester.ac.uk }

\date{%
	Department of Mathematics \\%
	The University of Manchester \\%
	Manchester, United Kingdom\\[2ex]%
}

\begin{document}

\maketitle

\begin{abstract}
  Mixture autoregressive (MAR) models provide a flexible way to model time
  series with predictive distributions which depend on the recent history of the
  process and are able to accommodate asymmetry and multimodality.  Bayesian
  inference for such models offers the additional advantage of incorporating the
  uncertainty in the estimated models into the predictions.  We introduce a new
  way of sampling from the posterior distribution of the parameters of MAR
  models which allows for covering the complete parameter space of the models,
  unlike previous approaches. We also propose a relabelling algorithm to deal a
  posteriori with label switching. We apply our new method to simulated and real
  datasets, discuss the accuracy and performance of our new method, as well as
  its advantages over previous studies.  The idea of density forecasting using
  MCMC output is also introduced.
\end{abstract}

\keyword{Mixture autoregressive model; 
		 Stationarity; 
		 MCMC methods; 
		 Model selection;  
		 Forecasting.}

\section{Introduction}

Mixture autoregressive (MAR) models \citep{WongLi2000} provide a flexible way to
model time series with predictive distributions which depend on the recent 
history of the process. Not only do the predictive distributions change over time,
they are also different for different horizons for predictions made at a fixed time point. 
As a consequence, they inherently accommodate asymmetry, multimodality and
heteroskedasticity. For this reason, mixture autoregressive models have been
considered a valuable alternative to other models for financial time series,
such as the SETAR model \citep{tong1990non}, the Gaussian transition mixture
distribution model \citep{LeEtAl1996},
or the widely used class of GARCH models \citep{nelson1991conditional}.

\citet{WongLi2000} considered estimation of MAR models based on the EM algorithm
\citep{dempster1977maximum}. That method is particularly well suited for
mixture-type models and works well.  On the other hand, a Bayesian approach can
offer the advantage of incorporating the uncertainty in the estimated models
into the predictions.

\citet{sampietro2006} presented the first Bayesian analysis of MAR
models. In his work, reversible jump MCMC \citep{green1995reversible} is used to select the 
autoregressive orders of the components in the mixture, and models with 
different number of components are compared using methods by \citet{Chib1995}
and \citet{ChibJeliazkov2001}, which exploit the marginal likelihood identity.
In addition, he derives analytically posterior distributions for 
all parameters in the selected model.

The Bayesian updates of the autoregressive parameters are problematic, because
the parameters need to be kept in the stationarity region, which is very
complex, and so cannot really be updated independently of each other.  In the
case of autoregressive (AR) models, it is routine to use parametrisation in
terms of partial autocorrelations \citep{Jones1987}, which are subject only to
the restriction to be in the interval $(-1,1)$.  \citet{sampietro2006} adapted
this neatly to MAR models by parameterising the autoregressive parameters of
each component of the MAR model with the partial autocorrelations of an AR model
with those parameters.

A major drawback of Sampietro's sampling algorithm for the autoregressive
parameters, is that it restricts the parameters of each component to be in the
stationarity region of an autoregressive model. While this guarantees that the
MAR model is stationary, it excludes from consideration considerable part of the
stationarity region of the MAR model (\citealp[p.~98]{WongLi2000};
\citealp{Boshnakov2011on1st2nd}). Depending on the mixture probabilities, the
excluded part can be substantial. For example,  most examples in
\citet[p.~98]{WongLi2000} cannot be handled by Sampietro's approach, see also
the examples in Section~\ref{sec:app}.

\citet{shahadat2012} developed a full analysis (model selection and sampling),
which reduced the constraints of Sampietro's analysis.  Using
Metropolis-Hastings algorithm and a truncated Gaussian proposal distribution for
the moves, he directly simulated the autoregressive parameters from their
posterior distribution.  This method still imposes a constraint on the
autoregressive parameters through the choice of boundaries for the truncated
Gaussian proposal. While the truncation is used to keep the parameters in the
stationarity region, the choice of boundaries is arbitrary and can leave out a
substantial part of the stationarity region of the model.  In addition, his
reversible jump move for the autoregressive order seems conservative, as it uses
functions which always prefer jumps towards low autoregressive orders (this will
be seen in Section \ref{sec:rjmcmc}).

A common problem associated with mixtures is label switching \citep[see for
instance][]{Celeux2000}, which derives from symmetry in the likelihood function.
If no prior information is available to distinguish components in the mixture,
then the posterior distribution will also be symmetric. It is essential that
label switching is detected and handled properly in order to obtain meaningful
results.  A common way to deal with this, also used by \citet{sampietro2006} and
\citet{shahadat2012}, is to impose identifiability constraints.  However, it is
well known that such constraints may lead to bias and other problems. In the
case of MAR models, \citet{shahadat2012} showed that these constraints may
affect convergence to the posterior distribution.

We develop a new procedure which resolves the above problems. 
We propose an alternative Metropolis-Hastings move to sample directly from the
posterior distribution of the autoregressive components. Our method covers the
complete parameter space.
We also propose a way of selecting optimal autoregressive orders using
reversible jump MCMC for choosing the autoregressive order of each component in
the mixture, which is less conservative than that of Hossain.
We propose the use of a
relabelling algorithm to deal a posteriori with label switching.

We apply the new method to both simulated and real datasets, and discuss the
accuracy and performance of our algorithm, as well as its advantages over
previous studies.
Finally, we briefly introduce the idea of density forecasting using MCMC output.

The structure of the paper is as follows. In Section~\ref{sec:mar-model} we
introduce the mixture autoregressive model and the notation we need.  In
Section~\ref{sec:analysis} we give detailed description of our method for
Bayesian analysis of MAR models, including model selection, full description of
the sampling algorithm, and the relabelling algorithm to deal with label
switching.  Section~\ref{sec:app} shows results from application of our method
to simulated and real dataset. Section~\ref{sec:prediction} introduces the idea
of density forecast using MCMC output.

\section{The mixture autoregressive model}
\label{sec:mar-model}

A process $\lbrace y_t \rbrace$ is said to follow a Mixture autoregressive (MAR)
process if its distribution function, conditional on past information, can be
written as
\begin{equation}\label{eq:MAR}
  F(y_t|\mathcal{F}_{t-1})
  = \sum_{k=1}^{g} \pi_k F_k\left(\dfrac{y_t - \phi_{k0} -
      \sum_{i=1}^{p_k}\phi_{ki}y_{t-i}}{\sigma_k}\right)
  ,
\end{equation}
where
\begin{itemize}

\item $\mathcal{F}_{t-1}$ is the sigma field generated by the process up to (and
  including) $t-1$. Informally, $\mathcal{F}_{t-1}$ denotes all the available
  information at time $t-1$, the most immediate past.

\item  $g$ is the total number of autoregressive components.

\item $\pi_k>0$, $k=1,\ldots,g$, are the mixing weights or proportions,
  specifying a discrete probability distribution. So, $\sum_{k=1}^{g}\pi_k = 1$
  and $\pi_g = 1 - \sum_{k=1}^{g-1}\pi_k$.  We will denote the vector of mixing
  weights by $\boldsymbol{\pi} = \left(\pi_1, \ldots, \pi_{g}\right)$.

\item $F_k$ is the distribution function (CDF) of a standardised distribution
  with location parameter zero and scale parameter one. 
  The corresponding density function will be denoted by $f_k$.

\item
  $\boldsymbol{\phi}_k = \left(\phi_{k0},\phi_{k1},\ldots,\phi_{kp_k}\right)$ is
  the vector of autoregressive parameters for the $k^{th}$ component, with 
  $\phi_{k0}$ being the shift. Here,
  $p_k$ is the autoregressive order of component $k$ and we define $p=\max(p_k)$ to be
  the largest order among the components. A useful convention is 
  to set $\phi_{kj} = 0$, for  $ p_k+1 \leq j \leq p$. 
  
\item $\sigma_k > 0 $ is the scale parameter for the $k^{th}$ component.  We
  denote by
  $\boldsymbol{\sigma} = \left(\sigma_1,\ldots\sigma_g\right)$ the vector of scale
  parameters. Furthermore, we define
  the precision, $\tau_k$,  of the $k^{th}$ component by
  $\tau_k = 1/\sigma_k^2$.
  
\item If the process starts at $t = 1$, then Equation~\eqref{eq:MAR} holds for
  $t>p$.

\end{itemize}
We will refer to the model defined by Equation~\eqref{eq:MAR} as
$\MAR(g;p_{1},\dots,p_{g})$ model.
The following notation will also be needed.
Let
\begin{equation*}
  \nu_{tk} = \phi_{k0} + \sum_{i=1}^{p_k} \phi_{ki}y_{t-i}
  .
\end{equation*}
The error term associated with  the $k$th component at time $t$ is defined by
\begin{equation}\label{eq:errterm}
  e_{tk}
  = y_t - \phi_{k0} - \sum_{i=1}^{p_k} \phi_{ki}y_{t-i}
  = y_t - \nu_{tk}
  .
\end{equation}
A useful alternative expression for $\nu_{tk}$ is the following mean corrected
form:
\begin{equation*}
  \nu_{tk} = \mu_k + \sum_{i=1}^{p_k} \phi_{ki}\left(y_{t-i}-\mu_k \right)
  .
\end{equation*}
Comparing the two representations we get
\begin{equation*}
  \phi_{k0} = \mu_k \bigl(1 - \sum_{i=1}^{p_k} \phi_{ki} \bigr)
  .
\end{equation*}
If $\sum_{i=1}^{p_k} \phi_{ki} \neq 0$, we also have
\begin{equation}\label{eq:transf1}
  \mu_k = \dfrac{\phi_{k0}}{1 - \sum_{i=1}^{p_k} \phi_{ki}}
  .
\end{equation}


A nice feature of this model is that the one-step predictive distributions
are given directly by the specification of the model with Equation~\eqref{eq:MAR}.
The $h$-steps ahead predictive distributions of $y_{t+h}$ at time~$t$ can be
obtained by simulation \citep{WongLi2000} or, in the case of Gaussian and
$\alpha$-stable components, analytically \citep{boshnakov2009mar}.

We focus here on mixtures of Gaussian components. In this case, using the
standard notations $\boldsymbol{\Phi}$ and $\boldsymbol{\phi}$ for the CDF and
PDF of the standard Normal distribution, we have $F_k \equiv \boldsymbol{\Phi}$
and $f_k \equiv \boldsymbol{\phi}$, for $k = 1,\ldots,g$.  The model in
Equation~\eqref{eq:MAR} can hence be written as
\begin{equation}\label{eq:MARgaussCDF}
  F(y_t|\mathcal{F}_{t-1})
  = \sum_{k=1}^{g} \pi_k \boldsymbol{\Phi}\left(
        \dfrac{y_t - \phi_{k0} - \sum_{i=1}^{p_k}\phi_{ki}y_{t-i}
             }{\sigma_k}
    \right)
\end{equation}
or, alternatively, in terms of the conditional  pdf
\begin{equation}\label{eq:MARgausspdf}
  f(y_t|\mathcal{F}_{t-1})
  = \sum_{k=1}^{g} \dfrac{\pi_k}{\sigma_k} \boldsymbol{\phi}\left(
        \dfrac{y_t - \phi_{k0} - \sum_{i=1}^{p_k}\phi_{ki}y_{t-i}
              }{\sigma_k}
      \right)
\end{equation}
Conditional mean and variance of $Y_{t}$ are 
\begin{equation} 
  \begin{split}
    \Ex[Y_t|\mathcal{F}_{t-1}]
    &= \sum_{k=1}^{g}\pi_k\left(\phi_{k0} + \sum_{i=1}^{p}\phi_{ki}y_{t-i}\right)
     =\sum_{k=1}^{g}\pi_k \mu_{tk}
    \\
    \Var(Y_t|\mathcal{F}_{t-1})
    &= \sum_{k=1}^{g}\pi_k \sigma^2_k
       + \sum_{k=1}^{g}\pi_k \mu_{tk}^2 - \sum_{k=1}^{g}\ \left(\pi_k \mu_{tk}\right)^2
  \end{split}
\end{equation}
The correlation structure of a MAR process with maximum order $p$ is similar to 
that of an $AR(p)$ process. At lag $h$ we have:
\begin{equation*}
  \rho_h = \sum_{k=1}^{g}\pi_k\sum_{i=1}^{p}\phi_{ki}\rho_{|h-i|}, \qquad h \geq 1
  .
\end{equation*}

\subsection{Stability of the MAR model}\label{sec:stability}

Stationarity conditions for MAR time series have some similarity to those for
autoregressions with some notable differences. Below we give the results we
need, see \citet{Boshnakov2011on1st2nd} and the references therein for further
details.

A matrix is stable if and only if all of its eigenvalues have moduli smaller
than one (equivalently, lie inside the unit circle).
Consider the companion matrices
\begin{equation*}
  A_k = \begin{bmatrix}
    &\phi_{k1} &\phi_{k2} &\dots &\phi_{k(p-1)} &\phi_{kp} \\
    &1 &0 &\dots &0 &0 \\
    &0 &1 &\dots &0 &0 \\
    &\vdots &\vdots &\ddots &\vdots &\vdots\\
    &0 &0 &\dots &1 &0 \\
  \end{bmatrix}
  , \quad
  k = 1,\ldots,g
  .
\end{equation*}
We say that the MAR model is stable if and only if the matrix.
\begin{equation*}
  A = \displaystyle \sum_{k=1}^{g} \pi_k A_k \otimes A_k
\end{equation*}
is stable ($\otimes$ is the Kronecker product).
If a MAR model is stable, then it can be used as a model for stationary time
series. The stability condition is  sometimes called stationarity condition. 

If $g = 1$, the MAR model reduces to an AR model and the above condition states
that the model is stable if and only if $A_1 \otimes A_1$ is stable, which is
equivalent to the same requirement for $A_1$.
For $g > 1$, it is still true that if all matrices $A_{k},\dots,A_{k}$,
$k=1,\dots,g$, are stable, then $A$ is also stable. However the inverse is no
longer true, i.e. $A$ may be stable even if one or more of the
matrices $A_{k}$ are not stable.

What the above means is that the parameters of some of the components of a
MAR model may not correspond to stationary AR models. It is
convenient to refer to such components as ``non-stationary''.

Partial autocorrelations are often used as parameters of autoregressive models
because they transform the stationarity region of the autoregressive parameters
to a hyper-cube with sides $(-1,1)$. The above discussion shows that the partial
autocorrelations corresponding to the components of a MAR model cannot be used
as parameters if coverage of the entire stationary region of the MAR model is
desired.


\section{Bayesian analysis of mixture autoregressive models}
\label{sec:analysis}
\subsection{Likelihood function and missing data formulation}

Given data $y_{1},\dots,y_{n}$, the likelihood function for the MAR model in the case of
Gaussian mixture components takes the form in (see Equation~\eqref{eq:MARgausspdf})
\begin{equation*}
  L(\boldsymbol{\phi},\boldsymbol{\sigma},\boldsymbol{\pi}|\boldsymbol{y})
  = \prod_{t=p+1}^{n} \sum_{k=1}^{g}
        \dfrac{\pi_k}{\sigma_k}
        \boldsymbol{\phi}
        \left(
          \dfrac{y_t - \phi_{k0} - \sum_{i=1}^{p_k} \phi_{ki}y_{t-i}}{\sigma_{k}}
        \right)
  .        
\end{equation*}
The likelihood function is not very tractable and a standard approach is to recur to
a missing data formulation \citep{dempster1977maximum}.

Let $\boldsymbol{Z}_t=\left(Z_{t1},\ldots,Z_{tg}\right)$ be a latent allocation random
variable, where $\boldsymbol{Z}_t$ is a g-dimensional vector
with entry $k$ equal to $1$ 
if $y_t$ comes from the $k^{th}$ component of the mixture,
and $0$ otherwise. We assume that the $\boldsymbol{Z_t}$s are discrete random
variables, independently drawn from the discrete distribution:
\begin{equation}
  P(Z_{tk} = 1|g,\boldsymbol{\pi}) = \pi_k
  , \qquad
  k=1,\ldots,g,
\end{equation}
This setup, widely exploited in the literature
\citep[see, for instance][]{dempster1977maximum,diebolt1994estimation} allows to
rewrite the likelihood function in a much more tractable way as follows:
\begin{equation}\label{eq:EMlik}
  L(\boldsymbol{\phi},\boldsymbol{\sigma},\boldsymbol{\pi}|\boldsymbol{y})
  = \prod_{t=p+1}^{n} \prod_{k=1}^{g} \left(
    \dfrac{\pi_k}{\sigma_k}
    \boldsymbol{\phi}\left(
      \dfrac{y_t - \phi_{k0} - 
                              \sum_{i=1}^{p_k} \phi_{ki}y_{t-i}}{\sigma_k}
    \right)\right)^{Z_{tk}} 
\end{equation}
In practice, the $\boldsymbol{Z}_t$s are not available.
We adopt a Bayesian approach to deal with this. We set suitable prior distributions on the
latent variables and the parameters of the model and develop a methodology for obtaining
posterior distributions of the parameters and dealing with other issues arising in the model
building process.

\subsection{Priors setup and choice of hyperparameters}

The setup of prior distributions is based on \citet{sampietro2006} and \citet{shahadat2012}.
In the absence of any relevant prior information it is natural to assume a priori that each
data point is equally likely to be generated from any component, i.e.
$\pi_{1} = \dots = \pi_{g} = 1/g$. This is a discrete uniform distribution, which is a
particular case of the multinomial distribution. The conjugate prior of the latter is the
Dirichlet distribution. We therefore set the prior for the mixing weigths vector,
$\boldsymbol{\pi}$, to
\begin{equation}
  \boldsymbol{\pi} \sim D\left(w_1,\ldots,w_g \right),
  \qquad w_1=\dots=w_g=1
  .
\end{equation}  
%
%
The prior distribution on the component means is a normal distribution with common fixed
hyperparameters $\zeta$ for the mean and $\kappa$ for the precision, i.e.
\begin{equation}
  \mu_k \sim N(\zeta, \kappa^{-1})
  , \qquad k=1,\ldots,g
  .
\end{equation}
For the component precisions, $\tau_k$, a hierarchical approach is adopted, as suggested in
\citet{RichardsonGreen1997}. Here, for a generic $k^{th}$ component the prior is a Gamma
distribution with hyperparameters $c$ (fixed) and $\lambda$, which itself follows a gamma
distribution with fixed hyperparameters $a$ and $b$. We have therefore
\begin{equation}
  \begin{split}
    c       &- \text{fixed} \\
    \lambda &\sim Ga(a, b)  \\
    \tau_k  &\sim Ga(c, \lambda)  ,\qquad k=1,\ldots,g
    .
  \end{split}
\end{equation}

The main difference between our approach and that of \citet{sampietro2006} and
\citet{shahadat2012} is in the treatment of the autoregressive parameters.

\citet{sampietro2006} exploits the one-to-one relationship between partial autocorrelations
and autoregressive parameters for autoregressive models descirbed in \citet{Jones1987}. 
Namely, he parameterises each MAR
component with partial autocorrelations, draws samples from the posterior distribution of the
partial autocorrelations via Gibbs-type moves and converts them to autoregressive parameters
using the functional relationship between partial autocorrelations and autoregressive
parameters. Of course, the term ``partial autocorrellations'' doesn't refer to the actual
partial autocorrellations of the MAR process, they are simply transformed parameters. 
The advantage of this procedure is that the stability region for the partial autocorrelation
parameters is just a hyper-cube with marginals in the interval $[-1,1]$, while for the AR
parameters it is a body whose boundary involves non-linear relationships between the
parameters.

A drawback of the partial autocorrelations approach in the MAR case is that it covers only
a subset of the stability region of the model. Depending on the other parameters, the loss
may be substantial. 

\citet{shahadat2012} overcomes the above drawbacks by 
simulating the AR parameters directly. He uses 
Random Walk Metropolis, while applying a constraint to the proposal distribution (a truncated
Normal). The truncation is chosen as a compromise that ensures that most of the stability
region is covered, while keeping a reasonable acceptance rate. Although effective with
"well behaved" data, there are scenarios, especially concerning financial examples,
in which the loss of information due to a pre-set truncation becomes significant, 
as will be shown later on.
In this paper, we choose Random Walk Metropolis for simulation from the
posterior distribution of autoregressive parameters, while exploiting the 
stability condition to avoid restraining the parameter space a priori.

With the above considerations, for the autoregressive parameters we choose a
multivariate uniform distribution with range
in the stability region of the model, and independence between parameters 
is assumed.
Hence, for a generic $\boldsymbol{\phi}_{k}$ 
 prior distribution is such that:
\begin{equation*}
  p(\boldsymbol{\phi}_{k})
  \propto \mathcal{I} \lbrace Stable \rbrace
  , \quad k = 1, \ldots, g
  .
\end{equation*} 
where $\mathcal{I}$ denotes the indicator function assuming value $1$ if the
condition is satisfied and $0$ otherwise.
We prefer this to a Normal prior as it better allows to explore the 
parameter space, and detect the presence of multimodality.


\paragraph{Choice of hyperparameters.}

Here we discuss the settings for the hyperparameters $\zeta$, $\kappa$, $a$, $b$, and $c$.
%
%
We have already discussed that the hyperparameters for the Dirichlet prior distribution
on the mixing weights (all equal to $1$). 
Also, $\lambda$ is a hyperparameter but it is a random variable with distribution which will
be fully specified once $a$ and $b$ are. 


Following \citet{RichardsonGreen1997}, let $\mathcal{R}_y= \max(y) - \min(y)$ be the length of the interval variation
of the dataset. Also fix the two hyperparameters $a=0.2$ and $c=2$. The remaining
hyperparameters are set as follows:

\begin{equation*}
\zeta = \min(y) + \dfrac{\mathcal{R}_y}{2} \qquad
\kappa = \dfrac{1}{\mathcal{R}_y}\qquad
b= \dfrac{100a}{c\mathcal{R}_y^2} = \dfrac{10}{\mathcal{R}_y^2}
\end{equation*}

\subsection{Posterior distributions and acceptance probability for RWM}

Following \citet{sampietro2006} and \citet{shahadat2012}, posterior distributions for all but the autoregressive parameters are as follows:
\begin{equation}
  \begin{split}
    P(z_{tk} = 1 \mid \boldsymbol{\pi}, \boldsymbol{\mu}, \boldsymbol{\tau}, \lambda,\boldsymbol{y})
    &=  \dfrac{ \pi_k\,\boldsymbol{\phi} \! \left(\dfrac{e_{tk}}{\sigma_k}\right)
              }{\displaystyle
                \sum_{l=1}^{g} \pi_l
                \boldsymbol{\phi}\left(\dfrac{e_{tl}}{\sigma_l}\right)
              }
    \\[5pt]
    \boldsymbol{\pi}  \mid \boldsymbol{\mu}, \boldsymbol{\phi},
                           \boldsymbol{\tau}, \boldsymbol{y}, \boldsymbol{z}
    &\sim D\left(1+n_1,\ldots,1+n_g\right)
    \\[5pt]
    \mu_k  \mid  \boldsymbol{\mu}_{-\mu_k}, \boldsymbol{\phi}, \boldsymbol{\tau},
              \boldsymbol{\pi}, \boldsymbol{y}, \boldsymbol{z}
    &\sim N \left(\dfrac{\tau_kn_k \bar{e}_k b_k + \kappa\zeta}{\tau_kn_kb_k^2+\kappa},
                  \dfrac{1}{\tau_kn_kb_k^2+\kappa}
            \right)
    \\[5pt]
    \lambda  \mid  \boldsymbol{\mu}, \boldsymbol{\phi}, \boldsymbol{\tau},
                \boldsymbol{\pi}, \boldsymbol{y}, \boldsymbol{z}
    &\sim Ga\left(a+gc,\, b + \displaystyle \sum_{k=1}^{g}\tau_k\right)
    \\[5pt]
    \tau_k  \mid  \boldsymbol{\mu}, \boldsymbol{\phi}, \boldsymbol{\tau}_{-\tau_k},
               \lambda, \boldsymbol{\pi}, \boldsymbol{y}, \boldsymbol{z}
    &\sim Ga\left(c + \dfrac{n_k}{2}, \,
                 \lambda + \dfrac{1}{2} \displaystyle \sum_{t=p+1}^{n} e_{tk}^2 z_{tk}
            \right)
  \end{split}
\end{equation}
where,  for $k=1, \ldots, g$, 
\begin{align*}
   e_{tk}     &= y_t - \nu_{tk} 
  ,&n_k       &= \sum_{t=p+1}^{n}z_{tk} 
  ,&b_k       &= 1 - \sum_{i=1}^{p_k}\phi_{ki} 
  ,&\bar{e}_k &= \dfrac{1}{n_k} \sum_{t=p+1}^{n} e_{tk} z_{tk}
  .                       
\end{align*}
All these parameters are updated via a Gibbs-type move.
Similarly, $\boldsymbol{Z}_t$s are simulated from a multinomial distribution
with associated posterior probabilities. 

To update autoregressive parameters, let $\boldsymbol{\phi}_{k}$,
$k=1, \ldots, g$, be the set of current states of the autoregressive
parameters, i.e. a set of
observations from the posterior distribution of $\boldsymbol{\phi}_{k}$. 
We can simulate 
 $\boldsymbol{\phi}_{k}^{*}$ from a
proposal $MVN(\boldsymbol{\phi}_{k},\Gamma_k^{-1})$ distribution, 
denoted by $q(\boldsymbol{\phi}_{k}^{*},\boldsymbol{\phi}_{k})$, 
with $\Gamma_k=\gamma_k I_{p_k}$, where $I_{p_k}$ is the identity 
matrix of size $p_k$. 

Here $\gamma_k$, $k=1,\ldots,g$ is a tuning parameter, chosen in such way that
the acceptance rate of RWM is optimal ($20-25\%$) for component $k$. We allow 
$\gamma_k$ to change between components, but to be constant within the same
component. 
Notice the difference between our proposal and the two-step approach by
 \citet{sampietro2006}, or the truncated Normal proposal chosen by \citet{shahadat2012}.
 The probability of accepting a move to the proposed $\boldsymbol{\phi}^{*}_k$ is
\begin{equation}
  \alpha\left(\boldsymbol{\phi}_{k}, \boldsymbol{\phi}_{k}^{*}\right)
  = \min \bigg \lbrace 1,
           \dfrac{f\left(\boldsymbol{y}  \mid  \boldsymbol{\phi}_k^{*} \right)
                  q\left(\boldsymbol{\phi}_k, \boldsymbol{\phi}_k^{*}\right)
                }{f\left(\boldsymbol{y}  \mid  \boldsymbol{\phi}_k\right)
                  q\left(\boldsymbol{\phi}_k^{*}, \boldsymbol{\phi}_k\right)
                }
         \bigg \rbrace
  ,
\end{equation} 
where $
  q\left(\boldsymbol{\phi}_k,    \boldsymbol{\phi}_k^{*}\right)
= q\left(\boldsymbol{\phi}_k^{*}, \boldsymbol{\phi}_k\right)
$, due to the symmetry in the Normal proposal. Therefore, the acceptance
probability will only depend on the likelihood ratio of the new set of
parameters over the current set of parameters, i.e.
\begin{equation}
  \alpha \left(\boldsymbol{\phi}_{k}, \boldsymbol{\phi}_{k}^{*}\right)
  =\min \bigg \lbrace
            1,
            \dfrac{f\left( \boldsymbol{y}  \mid  \boldsymbol{\phi}_k^{*}\right)
                 }{f\left( \boldsymbol{y} \mid \boldsymbol{\phi}_k\right) }
        \bigg \rbrace
\end{equation}
where
\begin{equation*}
  \dfrac{f\left(\boldsymbol{y}  \mid   \boldsymbol{\phi}_k^{*}\right)
       }{f\left(\boldsymbol{y} \mid \boldsymbol{\phi}_k\right)}
  = \dfrac{\displaystyle \prod_{\substack{t=p+1 \\ z_{tk}=1}}^{n} \exp \Bigg\lbrace 
        -\dfrac{1}{2\sigma_k^2} 
         \left(y_t-\phi^{*}_{k0} - \sum_{i=1}^{p_k}\phi^{*}_{ki}y_{t-i}\right)^2
    \Bigg\rbrace}{\displaystyle \prod_{\substack{t=p+1 \\ z_{tk}=1}}^{n} \exp \Bigg \lbrace
        -\dfrac{1}{2\sigma_k^2} 
         \left(y_t-\phi_{k0} -\sum_{i=1}^{p_k}\phi_{ki}y_{t-i}\right)^2
    \Bigg\rbrace}
\end{equation*}
This means that the likelihood ratio for the 
$k^{th}$ component is independent of current values of parameters
for the remaining components. This enables to calculate likelihood ratios separately
for each component.

The procedure described builds a candidate model with updated mixing weights,
shift, scale and autoregressive parameters. However, because stability of such 
model does not only depend on the autoregressive parameters, we must ensure that
the stability condition of Section~\ref{sec:stability} is satisfied. If this 
is not the case, the candidate model and all its parameters are 
rejected, and the current state of the chain is set to be the same as at the
previous iteration.

\subsection{Dealing with label switching}

Once the samples have been drawn, label switching is dealt with using a $k$-means clustering
algorithm proposed by \citet{Celeux2000}.
It is natural to use the identifiability constraint  $\pi_1>\pi_2>\dots>\pi_g$ but 
it is well known that it is problematic. Examples are given
in the discussion to the paper by \citet{RichardsonGreen1997}.
It was shown in fact by \citet[]{shahadat2012} that applying an identifiability 
constraint such as $\pi_1>\pi_2>\dots>\pi_g$ may in some cases affect convergence of the
chain. With our
approach instead, we do not 
interfere with the chain during the simulation, and hence convergence
is not affected.

Our algorithm works by first choosing the first $m$ simulated values of 
the output after convergence. The value $m$ shall be chosen small enough 
for labels switch 
to not have occurred yet, and large enough to be able to calculate reliable initial 
values of cluster centres and their respective variances.

Let $\boldsymbol{\theta}=\left(\theta_1,\ldots,\theta_g\right)$ be a subset of
 model parameters of size $q$, and $N$ the size of the converged sample. 
 For any centre coordinate $\theta_i$, $i=1,\ldots,q$
we calculate the mean and variance, based on the first $m$ simulated values, 
respectively as:
\begin{equation*}
\bar{\theta}_i=\dfrac{1}{m}\sum_{j=1}^{m}\theta^{(j)}_i \qquad 
\bar{s}^2_i=\dfrac{1}{m} \sum_{j=1}^{m}\left(\theta^{(j)}_i-\bar{\theta}_i\right)^2
\end{equation*}

We set this to be the ``true'' permutation of the components, i.e. we now have an initial center
$\boldsymbol{\bar{\theta}}^{(0)}$ with variances $\bar{s}^{(0)^2}_i$, $i=1,\ldots,q$. 
The remaining $g!-1$ permutations can be obtained by
simply permuting these centres.

From these initial estimates, the $r^{th}$ iteration ($r=1,\ldots,N-m$) of the
procedure consists of two steps:
\begin{itemize}
	\item the parameter vector $\boldsymbol{\theta}^{(m+r)}$ is assigned
	to the cluster such that the normalised squared distance
\begin{equation}
\sum_{i=1}^{g} \dfrac{\left(\theta^{(m+r)}_i - \bar{\theta}^{(m+r-1)}_{i}
	\right)^2}{\left(s^{(m+r-1)}_i\right)^2}
\end{equation}
is minimised,
where $\bar{\theta}^{(m+r-1)}_{i}$ is the $i^{th}$ centre coordinate 
and $s^{(m+r-1)}_i$ its standard deviation, at 
the latest update $m + r - 1$.
\item 
Centre coordinates and their variances are respectively updated as follows:
\begin{equation}
\bar{\theta}^{(m+r)}_i=\dfrac{m+r-1}{m+r}\bar{\theta}^{(m+r-1)}_i + \dfrac{1}{m+r}\theta^{(m+r)}_i
\end{equation}  
and
\begin{equation}
\begin{split}
  (s^{(m+r)}_i)^2
  &= \dfrac{m+r-1}{m+r}(s^{(m+r-1)}_i)^2 + \dfrac{m+r-1}{m+r}\left(
    \bar{\theta}^{(m+r-1)}_i-\bar{\theta}^{(m+r)}_i\right)^2
  \\ &\qquad {}
  + \dfrac{1}{m+r}\left( \theta^{(m+r)}_i-\bar{\theta}^{(m+r)}_i\right)^2 
\end{split}
\end{equation}
for $i=1,\ldots,q$.
\end{itemize}

For the mixture autoregressive case, it is not always clear which subset of the 
parameters should be used. In fact, group separation might seem clearer in the
mixing weights at times, as well as in the scale or shift parameters. 
Therefore this method requires graphical assistance, i.e. checking the raw
output looking for clear group separation.
However, it is advisable not to use the autoregressive parameters, especially 
when the orders are different.
 
Once the selected subset has been relabelled, labes for the remaining
parameters can be switched accordingly.

\subsection{Reversible Jump MCMC for choosing autoregressive orders}
\label{sec:rjmcmc}
For this step, we use Reversible Jump MCMC \citep[]{green1995reversible}. At each
iteration, one component $k$ is randomly chosen from the model. Let $p_k$ be the current
autoregressive order of this component, and set $p_{max}$ to be the largest possible
value $p_k$ may assume.  For the selected component, we propose to increase or decrease its
autoregressive order by $1$ with probabilities
\begin{equation*}
  p_k^{*} = \begin{cases}
    p_k - 1 & \text{with probability}~ d(p_k) \\
    p_k+1   & \text{with probability} ~b(p_k)
  \end{cases}
\end{equation*}  
where $b(p_k) = 1 - d(p_k)$, and such that $d(1) = 0$ and $b(p_{max}) = 0$. Notice that
$d(p_k)$ (or equivalently $b(p_k)$) may be any function defined in the interval $[0,1]$
satisfying such condition. For instance, \citet{shahadat2012} introduced two parametric functions for this
step. However, in absence of relevant prior information, we choose
$b(p_k) = d(p_k) = 0.5$ in our analysis, while presenting
the method in the general case.

Finally, it is necessary to point out that in both scenarios we have a 1-1 mapping
 between current and proposed model, 
so that the resulting Jacobian is always equal to $1$.

Given a proposed move, we proceed as follows:
\begin{itemize}

\item If the proposal is to move from $p_k$ to $p_k^{*}=p_k-1$, we simply drop $\phi_{kp_k}$,
  and calculate the acceptance probability by multiplying the likelihood ratio and the
  proposal ratio, i.e.
  \begin{multline}
    \alpha \left(\mathcal{M}_{p_k}, \mathcal{M}_{p_k^{*}}\right)
    \\ =
      \min \bigg \lbrace
        1, \dfrac{ f\left(\boldsymbol{y}  \mid  \boldsymbol{\phi}_k^{p_k^{*}} \right)
                }{ f\left(\boldsymbol{y}  \mid  \boldsymbol{\phi}_k^{p_k} \right)}
           \times
           \left[ \dfrac{b \left(p_k^{*} \right)}{ d\left(p_k\right)}
                  \times
                  \boldsymbol{\phi}\left(
                    \dfrac{\phi_{kp_k} - \phi_{kp_k}}{1/\sqrt{\gamma_k}}
                  \right)
           \right]
      \bigg \rbrace
  \end{multline}
  where
  $\boldsymbol{\phi}\left(\dfrac{\phi_{kp_k} -
      \phi_{kp_k}}{1/\sqrt{\gamma_k}}\right)$ is the density of the parameter
  dropped out of the model, according to its proposal distribution.

If the candidate model is not stable, then it is 
automatically rejected, i.e. $ \alpha \left(\mathcal{M}_{p_k}, \mathcal{M}_{p_k^{*}}\right)=0$.

\item If the proposed move is from $p_k$ to $p_k^{*} = p_k + 1$, we proceed by
  simulating the additional parameter from a suitable distribution. In absence
  of relevant prior information, the choice is to simulate a value from a
  uniform distribution centred in $0$ and with appropriate range, so that values both close and far apart
  from $0$, both positive and negative, are taken into consideration.
  
  These considerations lead to draw $
    \phi_{kp_k^{*}} \sim \mathcal{U}\left(-1.5, 1.5\right)$

  The acceptance probability is in this case   
  \begin{equation}\label{eq:pkplus1}
    \alpha\left(\mathcal{M}_{p_k}, \mathcal{M}_{p_k^{*}}\right)
    = \min \bigg \lbrace
        1, \dfrac{ f\left(\boldsymbol{y} \mid \boldsymbol{\phi}_k^{p_k^{*}}\right)
                }{ f\left(\boldsymbol{y} \mid \boldsymbol{\phi}_k^{p_k}\right)}
           \times
           \left[\dfrac{ d\left(p_k\right)}{b\left(p_k^{*}\right)} \times 3\right]
      \bigg \rbrace
  \end{equation}
  where $3$ is the inverse ofthe $\mathcal{U}\left(-1.5, 1.5\right)$ density.
  
  Once again, if the candidate model is not stable, $ \alpha \left(\mathcal{M}_{p_k},
   \mathcal{M}_{p_k^{*}}\right)=0$ and the current model is retained.
\end{itemize}

\subsection{Choosing the number of components}

To select the appropriate number of autoregressive components in the mixture, we
apply the methods proposed by \citet{Chib1995} and \citet{ChibJeliazkov2001},
respectively, for use of output from Gibbs and Metropolis-Hastings sampling. 
Both make use of the marginal likelihood identity.
	
From Bayes theorem,  we know that
\begin{equation}
  p(g | y) \propto f(y  \mid  g)p(g)
  ,
\end{equation}
where $p(g)$ is the prior distribution on $g$,  and $f(y \mid g)$ is the marginal likelihood function,  defined as
\begin{equation}
  f(y \mid g)
  = \sum_{p} \int f(y  \mid  \boldsymbol{\theta}, \boldsymbol{p, g})
                  p(\boldsymbol{\theta}, p  \mid  g)
                  d\boldsymbol{\theta}
\end{equation}
with
$\boldsymbol{\theta} = \left(
  \boldsymbol{\phi}, \boldsymbol{\pi}, \boldsymbol{\mu}, \boldsymbol{\tau}
\right)$ being the parameter vector of the model.
	
For any values $\boldsymbol{\theta^{*}}$, $p^{*}$, number of components $g$ and
observed data $\boldsymbol{y}$, we can use the marginal likelihood identity to
decompose the marginal likelihood into parts that are know or can be estimated
\begin{equation}\label{eq:likidentity}
  \begin{split}
  f(\boldsymbol{y}|g)
  &= \dfrac{f(\boldsymbol{y}  \mid  \boldsymbol{\theta^{*}}, p^{*}, g)
            p\left(\boldsymbol{\theta^{*}}, p^{*}  \mid  g \right)
          }{p\left(\boldsymbol{\theta^{*}}, p^{*}  \mid  \boldsymbol{y}, g\right)}
  \\ &=
     \dfrac{f(\boldsymbol{y}  \mid  \boldsymbol{\theta^{*}}, p^{*},g)
            p\left(\boldsymbol{\theta^{*}}  \mid  p^{*}, g\right)
            p(p^{*} \mid g)
          }{p\left(\boldsymbol{\theta^{*}}  \mid  p^{*}, \boldsymbol{y},g\right)
            p(p^{*}  \mid  \boldsymbol{y}, g)
          }
  \end{split}
\end{equation}
Notice that the only quantity not readily available in the above
equation 
is $p\left(\boldsymbol{\theta^{*}} \mid p^{*}, \boldsymbol{y}, g\right)$. However,
this can be estimated by running reduced MCMC simulations for fixed $p^{*}$
(which can be obtained by the RJMCMC method described in Section 5.1), as
follows:
\begin{equation}
    \hat{p}\left(\boldsymbol{\theta^{*}}  \mid  p^{*}, \boldsymbol{y}, g\right)
    =
    \begin{aligned}[t]
    & \hat{p}\left(\boldsymbol{\phi^{*}}  \mid  \boldsymbol{y}, p^{*}, g\right)
    \\ & \hat{p}\left(\boldsymbol{\mu^{*}}
                       \mid  \boldsymbol{\phi^{*}}, \boldsymbol{y}, p^{*}, g
                \right)
    \\ & \hat{p}\left(\boldsymbol{\tau^{*}}
                       \mid  \boldsymbol{\mu^{*}}, \boldsymbol{\phi^{*}},
                          \boldsymbol{y}, p^{*}, g
                 \right)
    \\ & \hat{p}\left(\boldsymbol{\pi^{*}}
                    \mid  \boldsymbol{\tau^{*}}, \boldsymbol{\mu^{*}},
                       \boldsymbol{\phi^{*}}, \boldsymbol{y}, p^{*}, g
                 \right)
    \end{aligned}
\end{equation}
Once these quantities are estimated (see \ref{eq:phihd},
\ref{eq:muhd}, \ref{eq:tauhd}, \ref{eq:pihd}), plug them in
Equation~\eqref{eq:likidentity}, together with the other known quantities, to
obtain the marginal likelihood for the model with fixed number of components
$g$.

For higher accuracy of results, it is suggested to compare marginal likelihood with different 
$g$ at points
of high density in the posterior distribution of $\boldsymbol{\theta}^{*}$. We will use
the estimated highest posterior density values.

\subsubsection*{Estimation of 
	$\hat{p}(\boldsymbol{\phi}^{*}  \mid  \boldsymbol{y}, p^{*}, g ) $}

Suppose we want to estimate
$\hat{p}\left(\boldsymbol{\phi_k^{*}}  \mid  p^{*}, \boldsymbol{y}, g\right)$, for
$k=1, \ldots, g$. We partition the parameter space into two subsets, namely
$\Psi_{k-1} = \left(p, \boldsymbol{\phi}_1, \ldots, \boldsymbol{\phi}_{k-1},
  g\right)$ and
$\Psi_{k+1} = \left(\boldsymbol{\phi}_{k+1}, \ldots, \boldsymbol{\phi}_{g},
  \boldsymbol{\mu}, \boldsymbol{\tau}, \boldsymbol{\pi}\right)$, where
parameters belonging to $\Psi_{k-1}$ are fixed (known or already selected
high density values).
	
First, produce a reduced chain of length $N_j$ to obtain
$\boldsymbol{\phi_k^{*}}$, the highest density value for $\boldsymbol{\phi_k}$,
using the sampling algorithm in Section 4.3, applied to the non-fixed set of 
parameters only. Define $\Psi_{k^{*}}$, the set of
known (fixed) parameters with the addition of $\boldsymbol{\phi_k^{*}}$.  From a
second reduced chain of length $N_i$, simulate
$\lbrace\tilde{\Psi}_{k+1}^{(i)}, \tilde{z}^{(i)}  \mid  \Psi_{k^{*}},
\boldsymbol{y}\rbrace$, as well as new observations
$\boldsymbol{\tilde{\phi}}_k^{(i)}$ from the proposal density in Equation 10,
centred in $\boldsymbol{\phi_k^{*}}$.
	
Now, let $\alpha(\boldsymbol{\phi}_k^{(j)}, \boldsymbol{\phi_k^{*}})$ and
$\alpha(\boldsymbol{\phi_k^{*}}, \boldsymbol{\tilde{\phi}}_k^{(i)})$ denote
acceptance probabilities respectively of the first and second chain.
We can finally estimate the value of the posterior density at
$\boldsymbol{\phi}_k^{*}$ as 
\begin{equation}
	\hat{p}\left(\boldsymbol{\phi}_k^{*}  \mid  p^\star, \boldsymbol{\phi}_1^{*}, \ldots, \boldsymbol{\phi}_{k-1}^{*}, g\right)=\dfrac{\dfrac{1}{N_j}\displaystyle \sum_{j=1}^{N_j}\alpha(\boldsymbol{\phi}_k^{(j)}, \boldsymbol{\phi_k^{*}}) q\left(\boldsymbol{\phi}_k^{(j)}, \boldsymbol{\phi}_k^{*}\right)}{\dfrac{1}{N_i}\displaystyle \sum_{i=1}^{N_i}\alpha(\boldsymbol{\phi_k^{*}}, \boldsymbol{\tilde{\phi}}_k^{(i)})}
\end{equation}
Repeat this procedure for all $k=1, \ldots, g$ and multiply the single densities to obtain
\begin{equation}\label{eq:phihd}
  \hat{p}\left(\boldsymbol{\phi}^{*}  \mid  \boldsymbol{y}, p^{*}, g \right)
  = \prod_{k=1}^{g}
      \hat{p}\left(
        \boldsymbol{\phi}_k^{*}
         \mid  p^\star, \boldsymbol{\phi}_1^{*}, \ldots, \boldsymbol{\phi}_{k-1}^{*}, g
      \right)
  .
\end{equation}
Note that there are no requirements on what $N_i$ and $N_j$ should be, granted
the first chain is long enough to have reached the stationary distribution.

\subsection*{Estimation of
  $\hat{p}\left(\boldsymbol{\mu}^{*}  \mid  \boldsymbol{\phi}^{*}, \boldsymbol{y}, p^{*}, g\right)$}

Run a reduced chain of length $N_i$. At each iteration, draw observations
$\boldsymbol{z}^{(i)}$, $\boldsymbol{\pi}^{(i)}$, $\boldsymbol{\tau}^{(i)}$,
$\boldsymbol{\mu}^{(i)}$. Set $\boldsymbol{\mu}^{*}=\left(\mu_1, \ldots, \mu_g\right)$, the
parameter vector of highest posterior density. The posterior density at
$\boldsymbol{\mu}^{*}$ can be estimated as
\begin{equation}\label{eq:muhd}
  \hat{p}\left(\boldsymbol{\mu}^{*}  \mid  \boldsymbol{\phi}^{*}, \boldsymbol{y}, p^{*}, g\right)
  =
  \dfrac{1}{N}\displaystyle\sum_{i=1}^{N}\prod_{k=1}^{g}
    p\left(
        \mu_k^{*}  \mid  \boldsymbol{\phi}^{*}, \boldsymbol{\tau}^{(i)},
        \boldsymbol{\pi}^{(i)}, \boldsymbol{y}, \boldsymbol{z}^{(i)}, p^{*}, g
      \right)
  .
\end{equation}
      
\subsection*{Estimation of
  $\hat{p}\left(\boldsymbol{\tau}^{*}  \mid  \boldsymbol{\mu}^{*}, \boldsymbol{\phi}^{*}, \boldsymbol{y}, p^{*}, g\right)$}

Run a reduced chain of length $N_i$. At each iteration, draw observations
$\boldsymbol{z}^{(i)}$, $\boldsymbol{\pi}^{(i)}$, $\boldsymbol{\tau}^{(i)}$. Set
$\boldsymbol{\tau}^{*}=\left(\tau_1, \ldots, \tau_g\right)$, the parameter
vector of highest posterior density. Posterior density at
$\boldsymbol{\tau}^{*}$ can be estimated as
\begin{equation}\label{eq:tauhd}
  \hat{p}\left(\boldsymbol{\tau}^{*}  \mid  \boldsymbol{\mu}^{*},
            \boldsymbol{\phi}^{*}, \boldsymbol{y}, p^{*}, g
         \right)
  = \dfrac{1}{N} \displaystyle \sum_{i=1}^{N} \prod_{k=1}^{g}
      p\left(
        \tau_k^{*}  \mid \boldsymbol{\mu}^{*}, \boldsymbol{\phi}^{*}, \boldsymbol{\pi}^{(i)},
        \boldsymbol{y}, \boldsymbol{z}^{(i)}, p^{*}, g
       \right)
  .
\end{equation} 

\subsection*{Estimation of 
	$\hat{p}\left(\boldsymbol{\pi}^{*}  \mid  \boldsymbol{\tau}^{*}, \boldsymbol{\mu}^{*}, \boldsymbol{\phi}^{*}, \boldsymbol{y}, p^{*}, g\right)$}

Run a reduced chain of length $N_i$. At each iteration, draw observations
$\boldsymbol{z}^{(i)}, \boldsymbol{\pi}^{(i)}$. Set
$\boldsymbol{\pi}^{*}=\left(\pi_1, \ldots, \pi_g\right)$, the parameter vector of highest
posterior density. Posterior density at $\boldsymbol{\pi}^{*}$ can be estimated as
\begin{equation}\label{eq:pihd}
  \hat{p}\left( \boldsymbol{\pi}^{*}  \mid  \boldsymbol{\tau}^{*}, \boldsymbol{\mu}^{*},
                \boldsymbol{\phi}^{*}, \boldsymbol{y}, p^{*}, g
         \right)
  = \dfrac{1}{N} \displaystyle \sum_{i=1}^{N} \prod_{k=1}^{g}
      p\left( \pi_k^{*}  \mid  \boldsymbol{y}, \boldsymbol{z}^{(i)}, p^{*}, g \right)
  .
\end{equation}

\section{Application}\label{sec:app}

For comparative and demonstrative purposes, we show applications of our method 
using two simulated datasets from \textbf{(A)} 
\begin{equation*}\footnotesize
  F(y_t|\mathcal{F}_{t-1})
  =  0.5\boldsymbol{\Phi}\left( \dfrac{y_t +0.5y_{t-1}}{1} \right)
   + 0.5\boldsymbol{\Phi}\left( \dfrac{y_t - y_{t-1}}{2}   \right)
\end{equation*}
and \textbf{(B)}
\begin{multline*}\footnotesize
  F(y_t|\mathcal{F}_{t-1})
  = 
     0.5\boldsymbol{\Phi}\left(\dfrac{y_t + 0.5y_{t-1} - 0.5y_{t-2}}{1}\right)
\\ \qquad {}
   + 0.3\boldsymbol{\Phi}\left(\dfrac{y_t + 0.4  y_{t-1}}{2}\right)
   + 0.2\boldsymbol{\Phi}\left(\dfrac{y_t - y_{t-1}}{4}\right)
 ,
\end{multline*}
respectively with $300$ and $600$ observations.
Process \textbf{(A)} is similar to the one considered by \citet{shahadat2012} and \citet{WongLi2000},
while \textbf{(B)} was chosen to illustrate in practice how labels switch
is dealt with. The issue of labels switch for \textbf{(B)} can be seen in Figure
\ref{fig:labswitch}, where we show
the raw MCMC output with signs of label switch between components 2 and 3 
(green and red lines), and the relabelled output after applying the algorithm.
\begin{figure}[!h]
  \centering
  \includegraphics[scale=0.4, width=10cm]{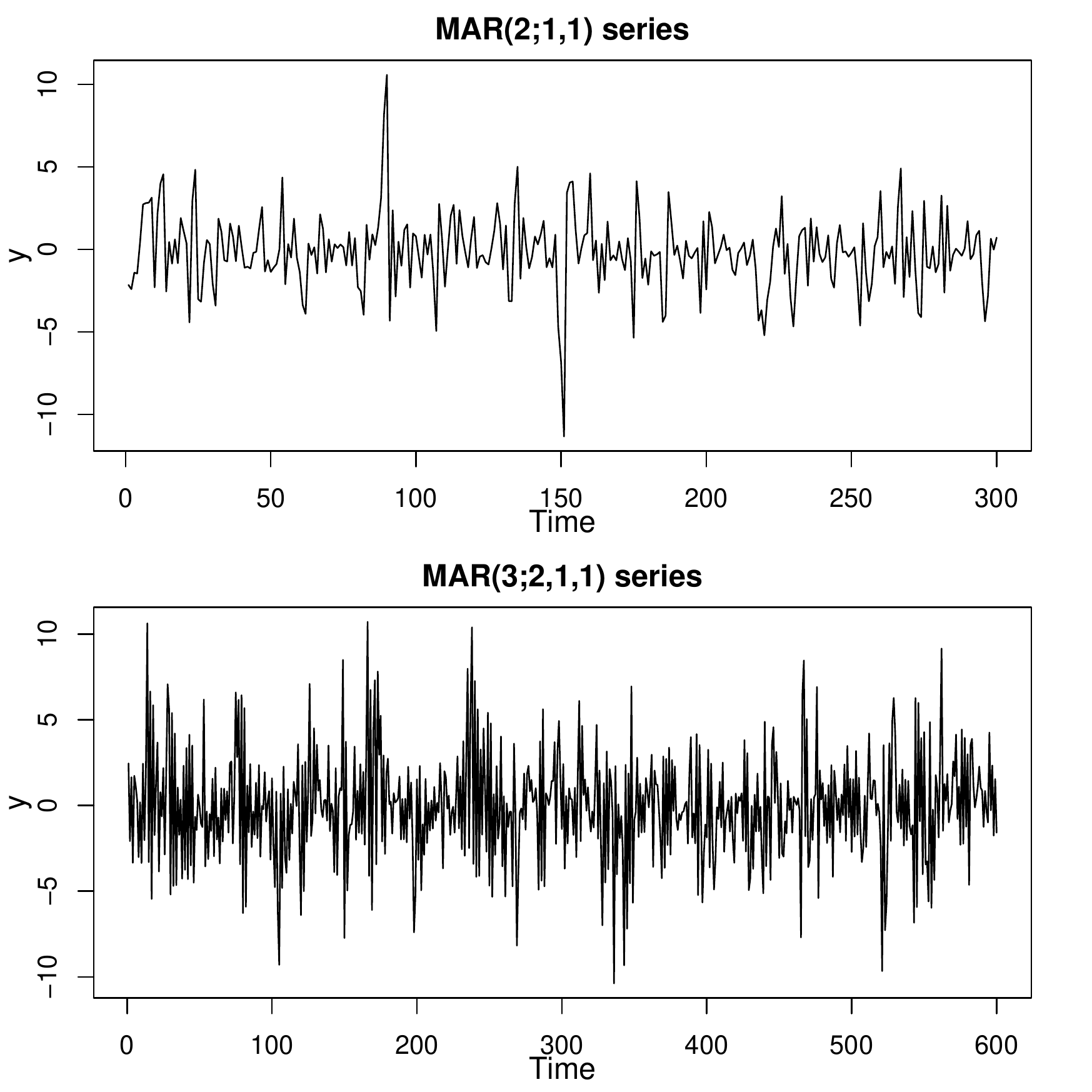}
  \caption{Simulated series from \textbf{(A)} (top) and \textbf{(B)} (bottom).}
\end{figure}

%
The algorithm then proceeds as described in Algorithm \ref{algorithm} below:
\begin{algorithm}
	\caption{}
	\begin{algorithmic}[1]
		\For{$g \gets 2,\ldots,g_{max}$}
		\State $\textit{RJMCMC and determine }p^{*}_1,\ldots,p^{*}_k$
		\State $\textit{Calculate }f(\boldsymbol{y}\mid g)$
		\EndFor
		\State $\textit{Select }g^{*} = \max f(\boldsymbol{y}\mid g)$, $g=2,\ldots,g_{max}$
		\State $\textit{Simulate }f\left(\boldsymbol{\theta}\mid y, g^{*}, \boldsymbol{p}^{*}\right)$
	\end{algorithmic}
\label{algorithm}
\end{algorithm}
\begin{table}[h]
  \footnotesize
  \parbox{.4\linewidth}{
    \centering
    \begin{tabular}{r|c|l}
      
      Model (A)&Preference & Marg. log-lik  \\
      \hline
      $\MAR(2;1, 1)$ &$0.7399$&$-611.8113$\\
      \hline
      $\MAR(3;1, 1, 1)$ &$0.1819$&$-613.0888$ \\
      \hline
      $\MAR(4;1, 1, 1, 4)$ &$0.0382$ &$-923.1585$
    \end{tabular}

  }
  \hfill
  \parbox{.4\linewidth}{
    \centering
    \begin{tabular}{r|c|l}
      Model (B)&Preference & Marg. log-lik  \\
      \hline
      $\MAR(2;2, 1)$ &$0.6258$&$-1468.628$\\
      \hline
      $\MAR(3;2, 1, 1)$ &$0.2937$&$-1383.061$ \\
      \hline
      $\MAR(4;2, 1, 2, 1)$ &$0.0491$ &$-1470.543$
    \end{tabular}

  }
  \caption{Results from simulation studies. ``Preference'' is the proportion of
    times the model was retained against all models with same number of
    components.}
 \label{tab:simrj}
\end{table}

\begin{figure}[h!]
  \centering
  \includegraphics[scale=.45]{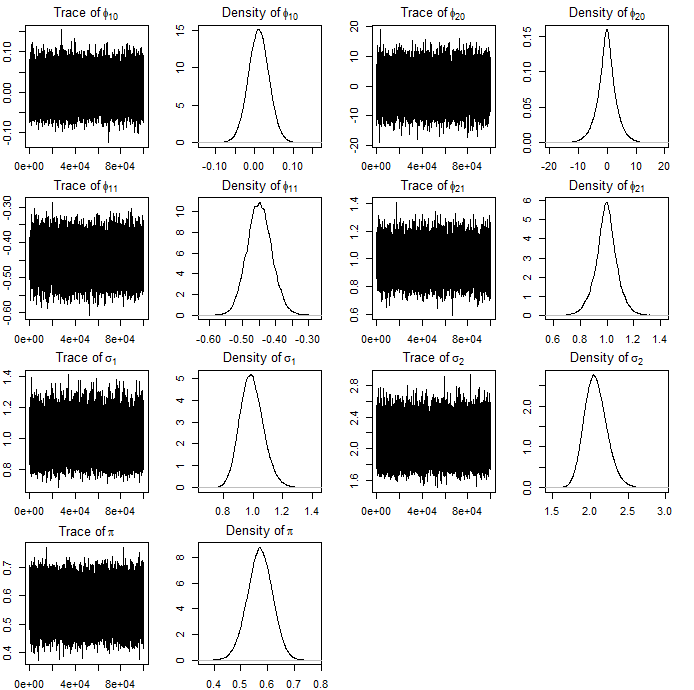}
  \caption{Trace and density plots of selected model from (A). Sample size is
    100000, after discarding 50000 observations as burn-in period.}
\label{fig:post1}
\end{figure}

\begin{figure}[!h]
  \centering
  \includegraphics[scale=0.3]{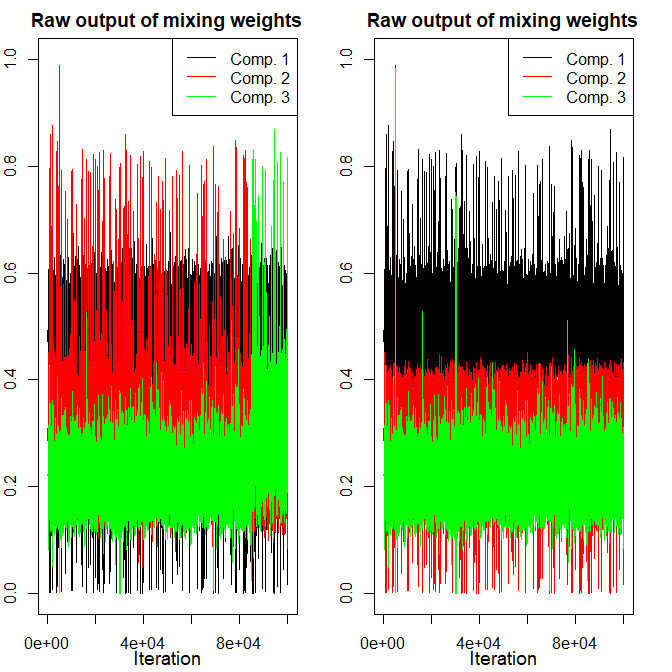}
  \caption{Comparison of raw output (left) and output adjusted for labels switch of mixing
    weights from \textbf{(B)}. We notice the effectiveness of the relabelling
    algorithm applied to our MCMC.}
\label{fig:labswitch}
\end{figure}

\begin{figure}[!h]
  \centering
  \includegraphics[scale=.62]{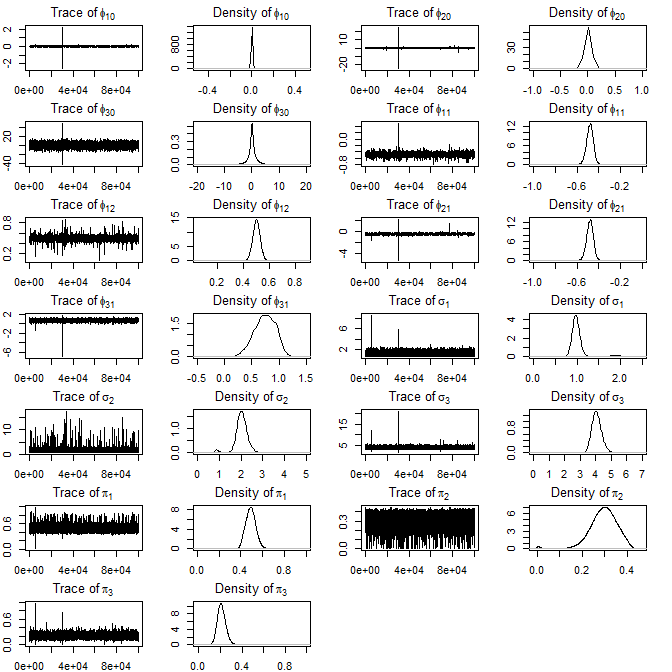}
  \caption{Trace and density plots of parameters from (B). Sample size is
    100000, after discarding 50000 observations as burn-in period.}
\label{fig:post2}
\end{figure}

As we can see from Tables, \ref{tab:simrj}, \ref{tab:post1} and \ref{tab:post2}, and Figures
\ref{fig:post1} and \ref{fig:post2}, the ``true'' model is chosen in both cases, as it has the
largest marginal log-likelihood.  In addition, true values of the parameters are found in high
density regions of their respective posterior distributions.
\begin{table}[!h]
  \centering
  \begin{tabular}{c|c|c|c|c}
    Model A&True Value &Posterior Mean &Standard Error &90\% HPDR \\
    \hline
    $\phi_{10}$ &0&0.011 &0.0268 &(-0.032,  0.055) \\
    \hline
    $\phi_{20}$ &0&-0.183     &3.273 &(-5.672,  5.206)\\
    \hline
    $\phi_{11}$ &-0.5&-0.449 & 0.037 &(-0.511,  -0.389) \\
    \hline
    $\phi_{21}$  &1&0.994 & 0.079 &(0.869,  1.136)  \\
    \hline
    $\sigma_1$ &1&0.992     & 0.079 &(0.862,  1.119)\\
    \hline
    $\sigma_2$ &2&2.069     &0.149 &(1.825,  2.311)\\
    \hline
    $\pi$ &0.5 &0.571    &0.046 & (0.494,  0.647)\\
  \end{tabular}
  \caption{Results of simulation from posterior distribution of the parameters
    under model (A).}
  \label{tab:post1}
\end{table}
      
\newpage

\begin{table}[!h]
  \centering
  \begin{tabular}{c|c|c|c|c}
    Model B&True Value &Posterior Mean &Standard Error &90\% HPDR \\
    \hline
    $\phi_{10}$ &0&0.001 &0.018 &(-0.009,  0.007) \\
    \hline
    $\phi_{20}$ &0&0.005     &0.253 &(-0.078,  0.091)\\
    \hline
    $\phi_{30}$ &0&0.102     &2.133 &(-3.145,  3.405)\\
    \hline
    $\phi_{11}$ &-0.5&-0.483 & 0.038 &(-0.536,  -0.427) \\
    \hline
    $\phi_{12}$ &0.5&0.498&0.034&(0.450,  0.547) \\
    \hline
    $\phi_{21}$  &-0.4&-0.461 & 0.105 &(-0.596,  -0.327)  \\
    \hline
    $\phi_{31}$ &1&0.731&0.264&(0.432,  1.058) \\
    \hline
    $\sigma_1$ &1&1.035    & 0.246 &(0.804,  1.156)\\
    \hline
    $\sigma_2$ &2&2.035     &0.439 &(1.625,  2.522)\\
    \hline
    $\sigma_3$ &4&4.074&0.341&(3.559,  4.573)\\		
    \hline
    $\pi_1$ &0.5 &0.495    &0.056 & (0.411,  0.568)\\
    \hline
    $\pi_2$ &0.3 &0.293    &0.064 & (0.207,  0.395)\\
    \hline
    $\pi_3$ &0.2 &0.212    &0.041 & (0.148,  0.275)
  \end{tabular}
  \caption{Results of simulation from posterior distribution of the parameters
    under model (B).}
  \label{tab:post2}
\end{table}

\newpage
     
To show consistency of the method, the experiment on model \textbf{(A)} was replicated several times. 
Details on that are available in the Appendix.  

\newpage 

\subsection{The IBM common stock closing prices}

The IBM common stock closing prices \citep{box2015time} 
is a financial time series widely explored several times
in the literature \citep[see, for instance][]{WongLi2000}. It contains 369 observations from
May $17^{th}~ 1961$ to November $2^{nd}~ 1962$.

\begin{figure}[!h]
  \centering
  \includegraphics[scale=.4, width=10cm]{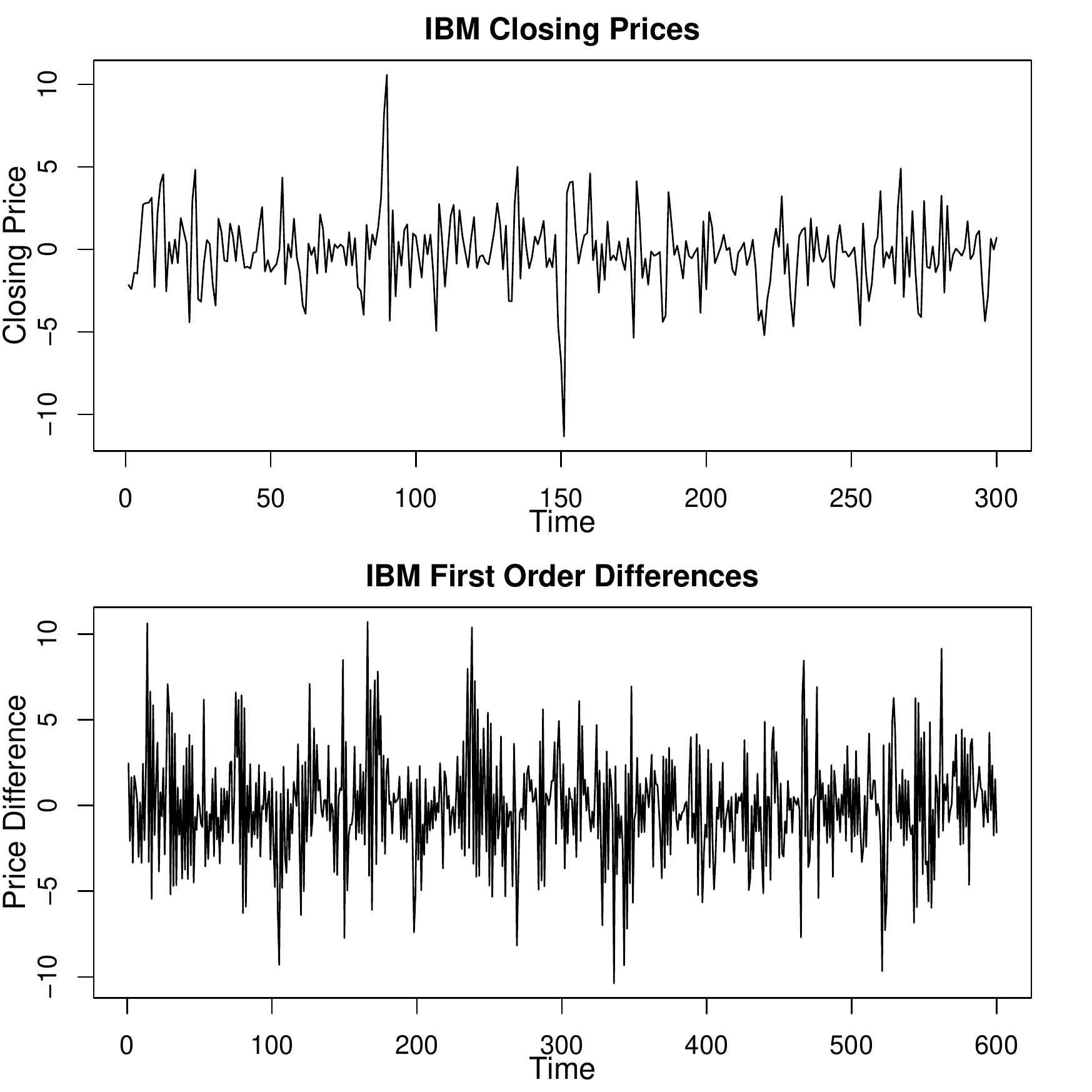}
  \caption{Times series of IBM closing prices (top) and series of the first
    order differences (bottom)}
\end{figure}

Following previous studies, we consider the series of first order differences. To
 allow direct comparison with \citet{WongLi2000} and \citet{shahadat2012}, we set $\phi_{k0}=0, ~k=1, 2, 3$.

\begin{figure}[h]
  \centering
  \includegraphics[scale=0.4, width=10cm]{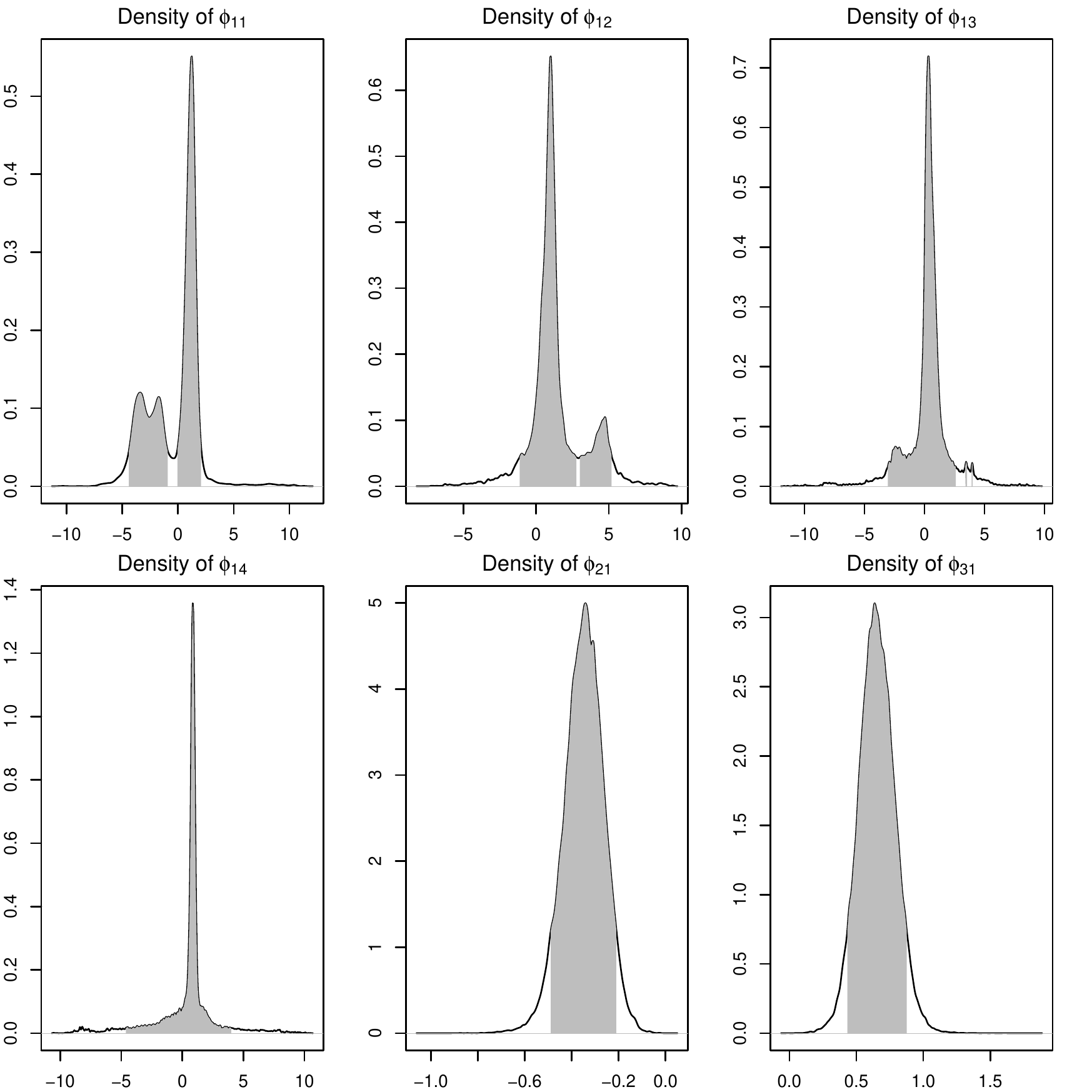}
  \caption{Posterior distributions of Autoregressive parameters from selected
    model $\MAR(3;4, 1, 1)$, with $90\%$ HPDR highlighted. We can clearly see
    multimodality occurring for certain parameters. Sample of $300000$ simulated
    values post burn-in.}
\label{fig:IBMpost}
\end{figure}

With the procedure outlined in Algorithm~\ref{algorithm} our method chooses a $\MAR(3;4, 1, 1)$ to best fit 
the data, amongst all 2, 3, and 4 component models of maximum order
$p_k=5, ~k=1, \ldots, g$, with a marginal log-likelihood of
$-1245.51$. We immediately notice that this is different from the selected model 
in \citet{WongLi2000}. Such difference may occur as the frequentist approach 
fails to capture the multimodality in the distribution of certain parameters, 
which we can clearly see from Figure \ref{fig:IBMpost}.
In fact, by attempting to fit a $\MAR(3;4, 1, 1)$ model by EM-Algorithm from 
several different starting points,
we concluded that this would actually provide a better fit than the
$\MAR(3;1, 1, 1)$ chosen by Wong and Li.

\newpage

\subsection{The Canadian lynx data}
Another dataset widely explored in time series literature, and in our interest by
\citet{WongLi2000}, is the annual record of Canadian lynx trapped in the Mackenzie
River district in Canada between 1821 and 1934. This dataset, listed by \citet{Elton1942TheTC}, 
includes $111$
observations.

\begin{figure}[!h]
\centering
\includegraphics[scale=0.5]{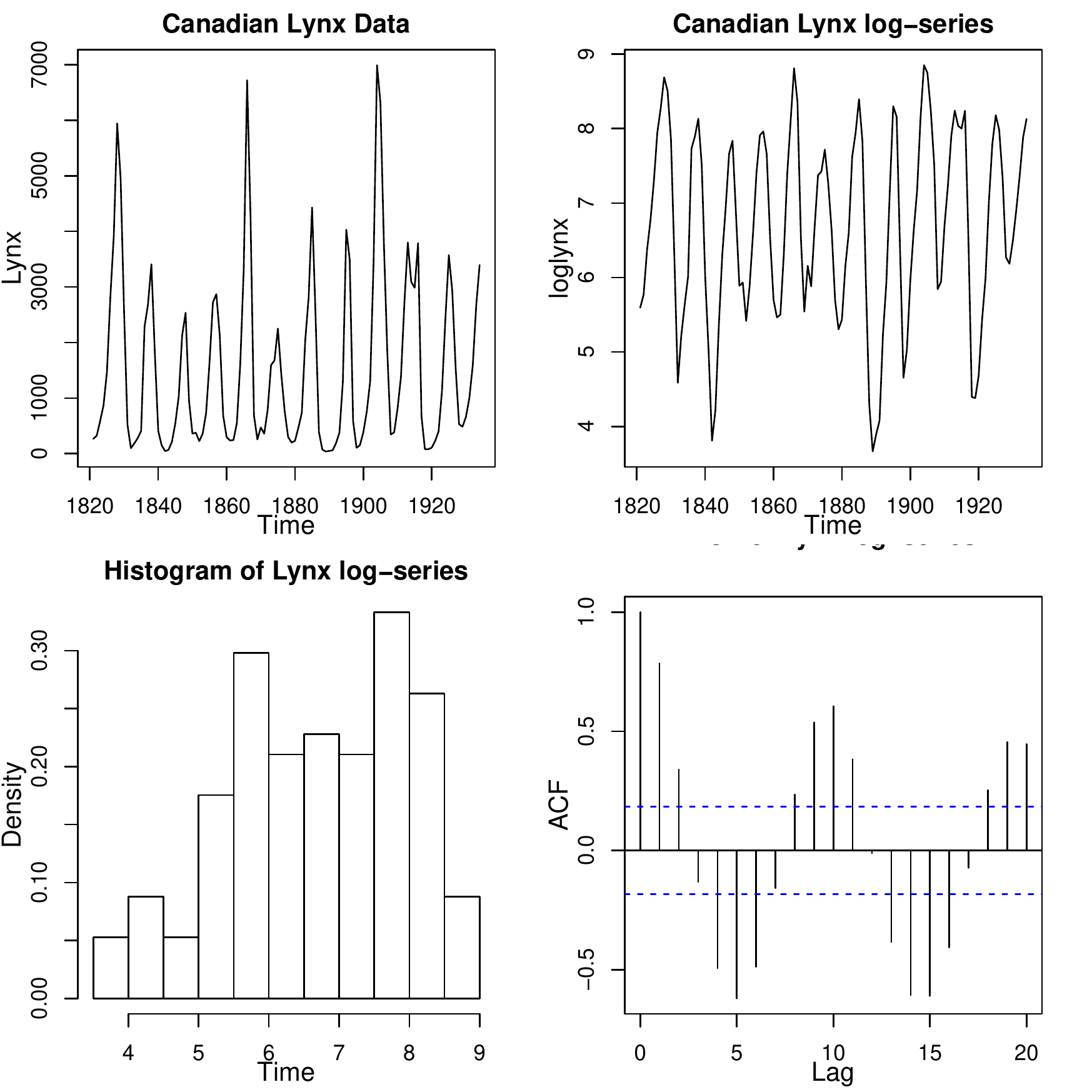}
\caption{Original time series of Canadian lynx (top left), series of natural 
	logarithms (top right), histogram of log-data (bottom left) and autocorrelation 
	plot of log-data (vottom right).
The data presents a typical autoregressive correlation structure, as well as 
multimodality.}
\label{fig:lynxplot}
\end{figure}

Following previous studies, we consider the natural logarithm
of the data, which presents a typical autoregressive correlation structure with 
$10$ years cycles. 
We notice the presence of multimodality in the log-data, with two local maxima (see Figure \ref{fig:lynxplot}).
This suggest that the series may be in fact generated by a mixture of two components.

In their analysis, \citet{WongLi2000} choose a $\MAR(2; 2, 2)$ as best model to fit
the data. However, their choice was based on  the minimum $BIC$ criterion, which
has been acknowledged for not always being reliable for MAR models, particularly
with small datasets. 

Aiming to have a better insight about the data, we apply 
our Bayesian method. The selected model is in this case a $\MAR(2; 1, 2)$, 
preferred over a $\MAR(2; 2, 2)$ by the algorithm, and to all $2, 3$ and $4$ component
models with autoregressive order $p=1,2,3,4$. The marginal log-likelihood of the
model is $-131.0381$. 
\begin{figure}
\centering
\includegraphics[scale=0.6]{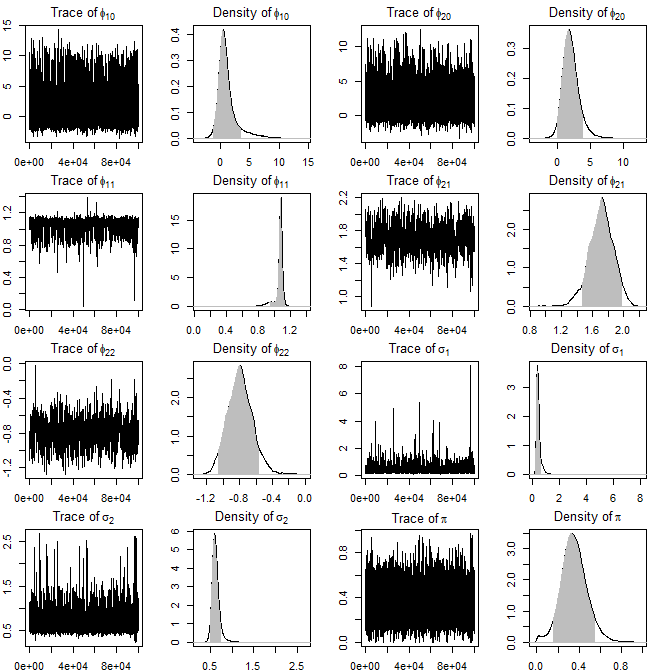}
\caption{Posterior trace plots and density of selected $\MAR(2;1,2)$ model for the
natural logarithm of Canadian lynx data. For all parameters, the credibility region
contains the estimated values from \citet{WongLi2000}.
Sample size is $100000$, after $50000$ burn-in iterations.}
\label{fig:lynxpost}
\end{figure}

We generated a sample of size $100000$ from the posterior 
distribution of the parameters of the selected $\MAR(2; 1, 2)$ model. It is noticed that,
for most paramters, the $90\%$ credibility region includes the MLEs obtained by
\citet{WongLi2000}. The only exception stands for the scale parameters, which seem
to be slightly larger than such MLEs. However, this may be due to our model containing one fewer AR parameter. 
On the other hand, these results are in line with the estimates
obtained by fitting a $\MAR(2;1,2)$ using the EM algorithm, since all estimates
are well within the corresponding $90\%$ highest posterior density region.

\begin{table}[!h]
	\centering
	\begin{tabular}{c|c|c|c|c}
		Parameter &MLE &HD value &Standard Error &90\% HPDR \\
		\hline
		$\phi_{10}$ &0.4957 &0.4962 &1.6897 &(-1.2599, 3.4341) \\
		\hline
		$\phi_{20}$ & 2.5728 &1.6945     &1.2663 &(-0.0138,  3.8897)\\
		\hline
		$\phi_{11}$ &0.9901 &1.0779 & 0.0667 &(0.9893 1.1320) \\
		\hline
		$\phi_{21}$  &1.5042 &1.7205 & 0.1594 &(1.4717,  1.9866)  \\
		\hline
		$\phi_{22}$ &-0.8984 &-0.7966 &0.1528 &(-1.0578, -0.5604) \\
		\hline
		$\sigma_1$ & 0.2313 &0.3553   & 0.1846 &(0.2162,  0.6451)\\
		\hline
		$\sigma_2$ &0.4828 &0.6010     &0.1006 &(0.4933,  0.7478)\\
		\hline
		$\pi$ &0.2358 &0.3280    &0.1247
		 & (0.1536,  0.5555)\\
	\end{tabular}
	\caption{Summary statistics of sample of size $100000$ from posterior distributions
		of the paramters of the selected model for the log-lynx data.}
	\label{tab:postlynx}
\end{table}
\newpage
\section{Bayesian density forecasts with mixture autoregressive models}
\label{sec:prediction}

Once a sample from the posterior is obtained,
it is useful to use these to make predictions on future (or off-set) observations.

\citet{WongLi2000} and \citet{boshnakov2009mar} respectively introduced a simulation based
and an analytical method for for density forecasts assuming a MAR model. 
The first method relies on Monte Carlo simulations, 
while the second derives exact h-step ahead predictive distributions of a given observation. 

On one hand, we could estimate density forecasts using the highest posterior density values (i.e. the peak of the posterior distribution). 
However, it is better in this case to exploit the entire simulated sample as follows:
\begin{enumerate}
\item Label each simulation from $1$ to $N$, e.g. $\boldsymbol{\theta}^{(i)}$, 
$i=1,\ldots,N$.
\item Calculate density forecast 
$f^{(i)}\left(y_{t+h} \mid \mathcal{F}_t, \boldsymbol{\theta}^{(i)} \right)$.
\item Estimate the density forecast
\begin{equation*}
\hat{f}\left(y_{t+h} \mid \mathcal{F}_t \right) =
\dfrac{1}{N} \sum_{i=1}^{N} f^{(i)}\left(y_{t+h} \mid \mathcal{F}_t, \boldsymbol{\theta}^{(i)} \right)
\end{equation*} 
\end{enumerate}
In this way, we obtain a sample from the h-steps ahead density forecast of an
observation of interest.

\begin{figure}[!htb]\centering
	\begin{minipage}{0.48\textwidth}
		\includegraphics[width=1\linewidth]{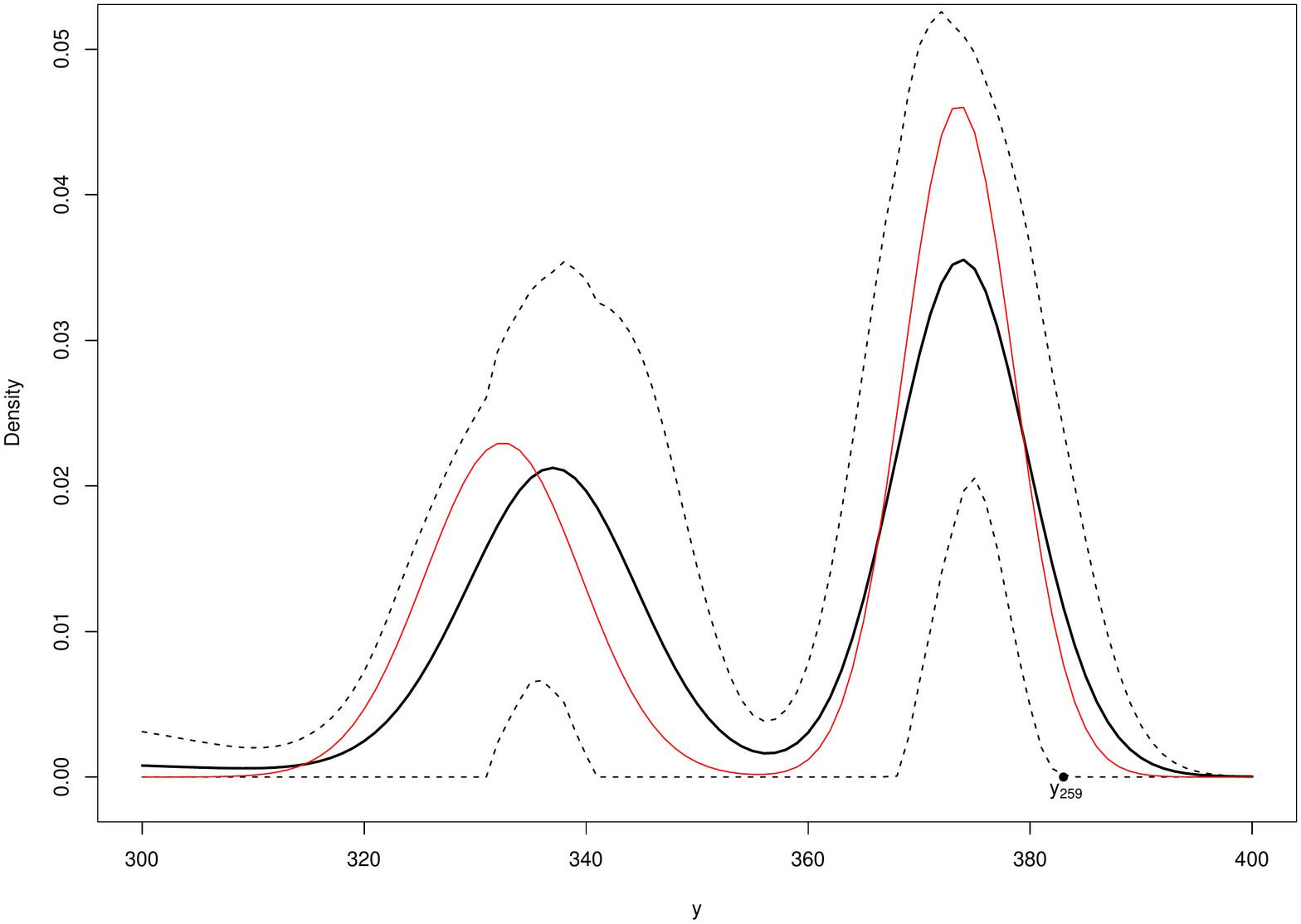}
	\end{minipage}
	\begin {minipage}{0.48\textwidth}
	\includegraphics[width=1\linewidth]{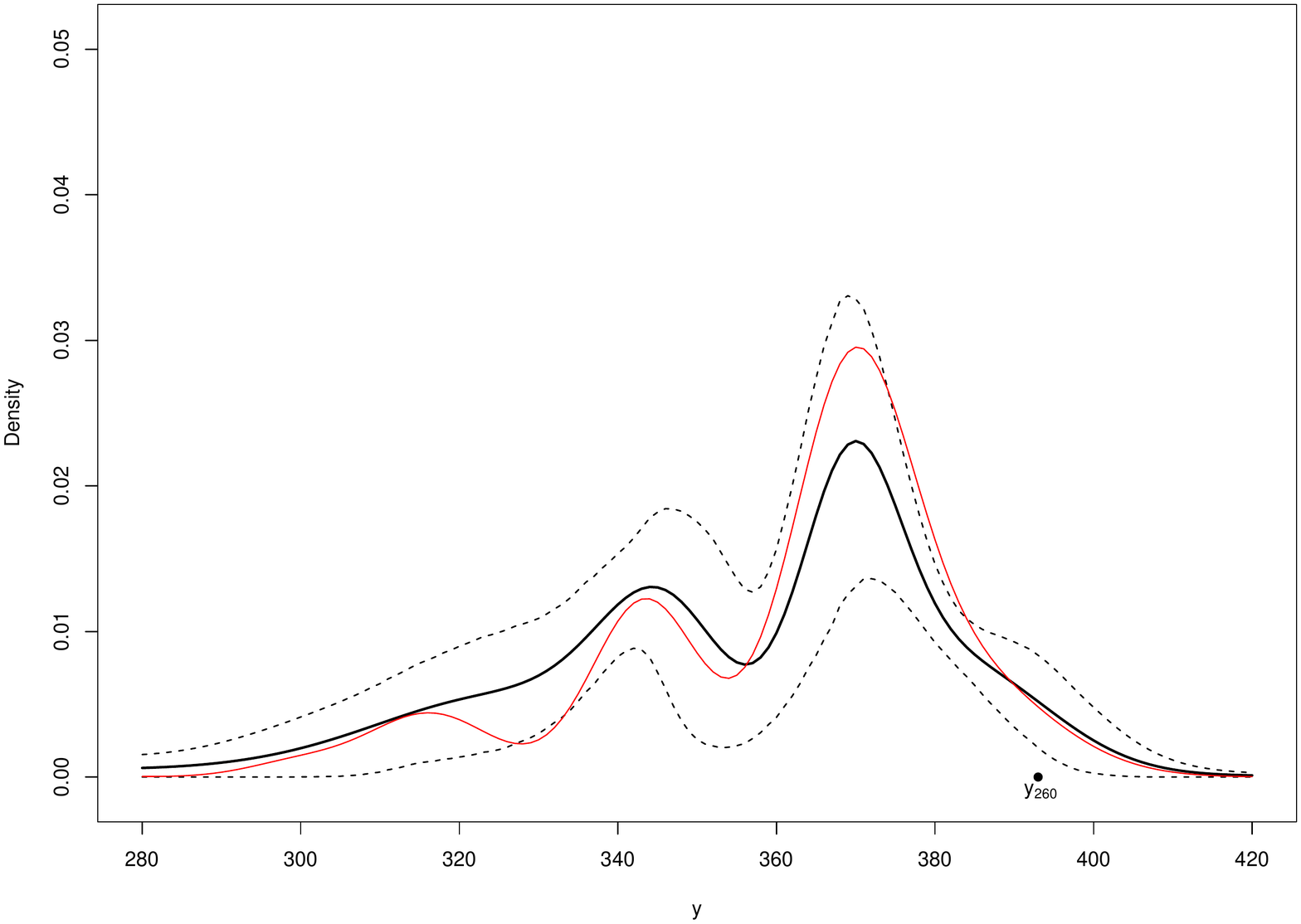}
\end{minipage}
\caption{Density of 1 and 2 steps ahead predictor at $t=258$ for the IBM data.
The solid black line represents our Bayesian method, with the 90\% credible 
interval identified by the dashed lines. The solid red line represents 
the predicted density using parameter values from EM estimation by Wong and Li.}
\label{fig:y258pred}
\end{figure}

We estimate the 1-step and
2-steps predictive distributions of the IBM data at $t=258$ using the analytical method by
\citet{boshnakov2009mar}, and compare them to the ones obtained by EM algorithm.
(see Figure \ref{fig:y258pred}). The solid red lines represent the density obtained by
\citet{boshnakov2009mar} using EM estimates and the exact method.
Results of our method are represented by the solid black lines, with the dashed lines 
as $90\%$ credibility region.
The figure also shows how quickly the uncertainty on the predictions grows as we
move further in the future, with the 2-step predictive density looking
much flatter. 

We can see that there are no substantial differences in the shape
of these predictive distributions. However, we notice that, particularly for the
2-steps predictor, averaging seems to "stabilise" the density line. 

We notice from the plots that, clearly for the 1-step predictor and
slighlty for the 2-step predictor, the density obtained by MCMC attaches 
higher density the observations of interest $y_{259}$ and $y_{260}$.

\section{Conclusion}

We presented an innovative fully Bayesian analysis of mixture autoregressive
models with Gaussian components, in particular a new method for simulation from
the posterior distribution of the autoregressive parameters, which covers the
whole stationarity region, compared to previous approaches that constrained it
in one way or another.  Our approach allowed us to better capture presence of
multimodality in the posterior distribution of model parameters. 
We also introduced a way of dealing with label switching that
does not interfere with convergence to the posterior distribution
of the model parameters. This consisted in using a
relabelling algorithm a posteriori. 

Simulations indicate that the method works well. We presented results
for two simulated data sets. In both
cases the ``true'' model was selected, and posterior distributions
showed high densities regions around the ``true'' values of the
parameters.

The ability of our method to explore the complete stationarity region of the
autoregressive parameters allows it to capture better multimodality of
distributions. This was illustrated with the IBM and the Canadian
Lynx datasets.
In the former (Figure \ref{fig:IBMpost}) we saw how multimodality in the
posterior distribution of 
autoregressive parameters was captured, aspects which were missed in the analyses of \citet{shahadat2012}, (see for example Figures 3.10 and 3.11).
 For this example, it was also noticed
that modes of posterior distributions of the autoregressive parameters roughly
 correspond to point estimates obtained by EM
estimation. 
In the latter (Figure \ref{fig:lynxpost}), we found the mode of $\phi_{21}$ to 
be quite distant from $0$, with values close to $2$ lying in the credibility interval. 
In this case, the risk with Hossain's method would be to truncate the Normal
proposal at points such that a significant part of the stationarity region
of the model is not covered. Sampietro's
method would have failed to detect such a mode, since it is 
outside the interval $[-1,1]$.

In conclusion, we may say that our algorithm provides accurate and informative
estimation, and therefore may result in more accurate predictions.

Further work could be done to improve the efficiency of our method. 
Possible improvements
to the method include a different algorithm for sampling of autoregressive
parameters.

In particular, acceptance rates for the Random Walk Metropolis moves used for
sampling the autoregressive parameters can be rather low for mixtures of large
number of components or for components with large autoregressive orders,
making the algorithm slow at times, with the added risk of it not
being able to explore the complete parameter space efficiently.  A different
procedure, such as the Metropolis Adjusted Langevin Algorithm (MALA),
may be considered to improve the efficiency.

Gaussian mixtures are very flexible but alternatives are worth considering.
In particular, components with standardised t-distribution could allow modelling
heavier tails with small number of components. 
\newpage
\appendix

\section*{Appendix}

We explain here how consistency of the method was assessed, with application to
data generated from Model \textbf{(A)} in Section \ref{sec:app}.

For this experiment, we simulate $400$ different datasets of length $n=300$ from
the underlying MAR process in Model \textbf{(A)}, and proceeded as follows:
\begin{enumerate}

\item For each dataset, we simulate a sample of size $100000$ from the posterior
distribution of the parameters, after allowing $10000$ iterations as burn-in period.

\item For each parameter, we find the overall minimum and maximum over the $400$ 
samples, say $l$ and $u$. 
From here, we identify a grid of $512$ equally spaced values in the range $[l,u]$,
and evaluate the density of such points under each posterior.

\item Finally, we average for each of the points to obtain a unique average density.
 
\end{enumerate}

The figure below summarises results of applying this procedure. As we can see, the
densities are well in line with the true values of the parameters.

\begin{figure}[!h]
\centering
\includegraphics[scale=0.5]{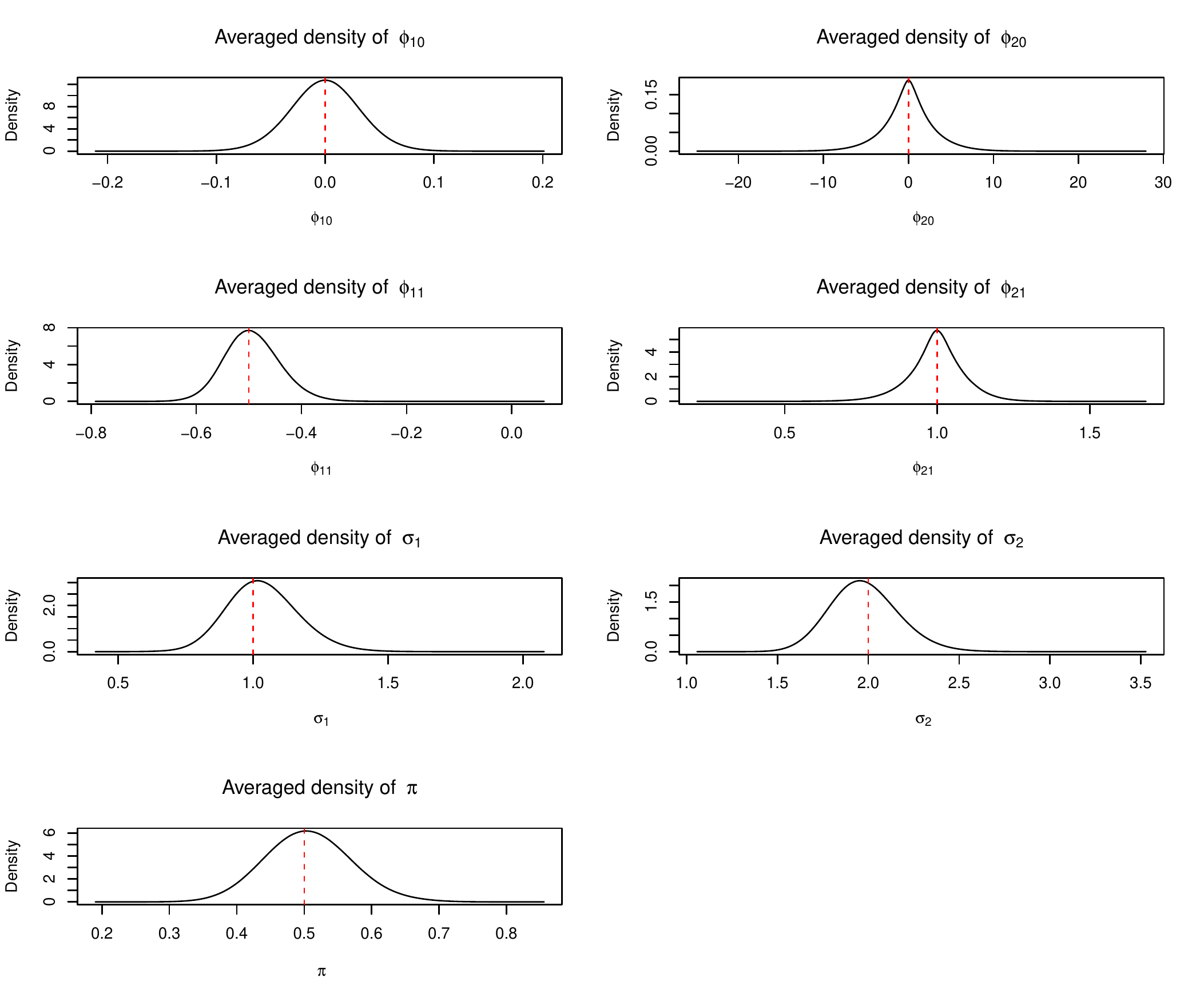}
\caption{Average densities of the parameters over $400$ simulated datasets of
	length $n=300$. Each simulation is a sample of size $100000$ from the posterior
distribution of the parameters.}
\end{figure}

\newpage
\bibliographystyle{kluwer}
\bibliography{marbayes}

@article{dempster1977maximum,
	title = {Maximum likelihood from incomplete data via the EM algorithm},
	author = {Dempster, Arthur P and Laird, Nan M and Rubin, Donald B},
	journal = {Journal of the royal statistical society. Series B (methodological)},
	pages = {1--38},
	year = {1977},
	publisher = {JSTOR}
}

@article{diebolt1994estimation,
  title = {Estimation of finite mixture distributions through Bayesian sampling},
  author = {Diebolt, Jean and Robert, Christian P},
  journal = {Journal of the Royal Statistical Society. Series B (Methodological)},
	volume  =  {56},
  pages = {363--375},
  year = {1994},
  publisher = {JSTOR}
}

@article{green1995reversible,
	title = {Reversible jump Markov chain Monte Carlo computation and Bayesian model determination},
	author = {Green, Peter J},
	journal = {Biometrika},
	volume = {82},
	number = {4},
	pages = {711--732},
	year = {1995},
	publisher = {Oxford University Press}
}

@article{nelson1991conditional,
	title = {Conditional heteroskedasticity in asset returns: A new approach},
	author = {Nelson, Daniel B},
	journal = {Econometrica: Journal of the Econometric Society},
	pages = {347--370},
	year = {1991},
	publisher = {JSTOR}
}

@book{tong1990non,
	title = {Non-linear time series: a dynamical system approach},
	author = {Tong, Howell},
	year = {1990},
	publisher = {Oxford University Press}
}

@article{Boshnakov2011on1st2nd,
AUTHOR  =  {Boshnakov, Georgi N.},
TITLE  =  {On First and Second Order Stationarity of Random Coefficient Models},
   JOURNAL  =  {Linear Algebra Appl.},
  FJOURNAL  =  {Linear Algebra and its Applications},
    VOLUME  =  {434},
      YEAR  =  {2011},
    NUMBER  =  {2},
     PAGES  =  {415--423},
  doi  =  {10.1016/j.laa.2010.09.023},
}

@article{boshnakov2009mar,
author = "Boshnakov, Georgi N.",
title = "{Analytic expressions for predictive distributions in mixture
    autoregressive models.}",
journal = "Stat. Probab. Lett. ",
volume = "79",
number = "15",
pages = "1704-1709",
year = "2009",
doi = {10.1016/j.spl.2009.04.009},
}

@article{sampietro2006,
  title = {Bayesian analysis of mixture of autoregressive components with an application to financial market volatility},
  author = {Sampietro, S.},
  journal = {Applied Stochastic Models in Business and Industry},
  volume = {22},
  number = {3},
  pages = {242},
  year = {2006},
  publisher = {John Wiley and Sons Ltd.}
}

@article{LeEtAl1996,
author = "Le, Nhu D. and Martin, R.Douglas and Raftery, Adrian E.",
title = "{Modeling flat stretches, bursts, and outliers in time series using
    mixture transition distribution models.}",
journal = "J. Am. Stat. Assoc. ",
volume = "91",
number = "436",
pages = "1504-1515",
year = "1996",
doi = {10.2307/2291576},
}

@article{WongLi2000,
author = "Wong, C. S. and  Li, W. K.",
title = "{On a mixture autoregressive model.}",
journal = "J. R. Stat. Soc., Ser. B, Stat. Methodol. ",
volume = 62,
number = 1,
pages = "95-115",
year = 2000,
}

@article{RichardsonGreen1997,
author = "Richardson, S. and  Green, P. J.",
title = "{On Bayesian Analysis of Mixtures with an Unknown Number of Components.}",
journal = "J. R. Stat. Soc., Ser. B, Stat. Methodol. ",
volume = "59",
number = "4",
pages = "731-792",
year = "1997",
}

@book{Celeux2000,
author = "Celeux, G.",
title = "{Bayesian Inference of Mixture: The Label Switching Problem.}",
publisher = "Payne R., Green P. (eds) COMPSTAT. Physica, Heidelberg",
pages = "227-232",
year = "2000",
}

@article{ChibJeliazkov2001,
author = "Chib, S. and  Jeliazkov, I.",
title = "{Marginal likelihood from the Metropolis-Hastings output.}",
journal = "J. A. Stat. Ass.",
volume = "96",
number = "453",
pages = "270-281",
year = "2001",
}

@article{Chib1995,
author = "Chib, S.",
title = "{Marginal likelihood from the Gibbs output.}",
journal = "J. A. Stat. Ass.",
volume = "90",
number = "432",
pages = "1313-1321",
year = "1995",
}

@PhdThesis{shahadat2012,
  author  =  {A.B.M. Shahadat Hossain},
  title   =  {Complete {Bayesian} analysis of some mixture time series models},
  school  =  {Probability and Statistics Group, School of Mathematics, University of
                Manchester},
  year    =  {2012},
}

@article{Elton1942TheTC,
	ISSN = {00218790, 13652656},
	URL = {http://www.jstor.org/stable/1358},
	author = {Charles Elton and Mary Nicholson},
	journal = {Journal of Animal Ecology},
	number = {2},
	pages = {215--244},
	publisher = {[Wiley, British Ecological Society]},
	title = {The Ten-Year Cycle in Numbers of the Lynx in Canada},
	volume = {11},
	year = {1942}
}

@article{Jones1987,
	ISSN = {00359254, 14679876},
	URL = {http://www.jstor.org/stable/2347544},
	author = {M. C. Jones},
	journal = {Journal of the Royal Statistical Society. Series C (Applied Statistics)},
	number = {2},
	pages = {134--138},
	publisher = {[Wiley, Royal Statistical Society]},
	title = {Randomly Choosing Parameters from the Stationarity and Invertibility Region of Autoregressive-Moving Average Models},
	volume = {36},
	year = {1987}
}

@Book{ box2015time,
	author = { Box, George E. P. and Jenkins, Gwilym M. },
	title = { Time series analysis : forecasting and control / George E.P. Box and Gwilym M. Jenkins },
	edition = { Rev. ed. },
	isbn = { 0816211043 },
	publisher = { Holden-Day San Francisco },
	pages = { xxi, 575 p. : },
	year = { 1976 },
	type = { Book },
	language = { English },
	subjects = { Time-series analysis.; Prediction theory.; Transfer functions.; Feedback control systems -- Mathematical models. },
	life-dates = { 1976 -  },
	catalogue-url = { https://nla.gov.au/nla.cat-vn640184 },
}

\end{document}